\documentclass[sigconf]{acmart}
\pdfoutput=1
\AtBeginDocument{%
  }

\copyrightyear{2024} 
\acmYear{2024} 
\setcopyright{acmlicensed}\acmConference[MM '24]{Proceedings of the 32nd
ACM International Conference on Multimedia}{October 28-November 1,
2024}{Melbourne, VIC, Australia}
\acmBooktitle{Proceedings of the 32nd ACM International Conference on
Multimedia (MM '24), October 28-November 1, 2024, Melbourne, VIC, Australia}
\acmDOI{10.1145/3664647.3680977}
\acmISBN{979-8-4007-0686-8/24/10}

\usepackage{multirow} 
\usepackage{graphicx}
\usepackage{color} 
\usepackage{fancyhdr}
\usepackage{hyperref}
\usepackage[normalem]{ulem} 

\begin{document}

\title{Navigating Weight Prediction with Diet Diary}


\author{Yinxuan Gui}
\orcid{0009-0004-9345-3275}
\affiliation{%
  \institution{Shanghai Key Lab of Intell. Info. Processing, School of CS, Fudan University, China
}
  \city{}
  \country{}}
\email{22210240154@m.fudan.edu.cn}

\author{Bin Zhu}
\orcid{0000-0002-9213-2611}
\affiliation{%
  \institution{Singapore Management University}
  \city{}
  \country{Singapore}
}
\email{binzhu@smu.edu.sg}

\author{Jingjing Chen}
\authornote{Corresponding author.}
\orcid{0000-0003-1737-3420}
\affiliation{%
 \institution{Shanghai Key Lab of Intell. Info. Processing, School of CS, Fudan University, China}
\city{}
\country{}}
\email{chenjingjing@fudan.edu.cn}

\author{Chong-Wah Ngo}
\orcid{0000-0003-4182-8261}
\affiliation{%
  \institution{Singapore Management University}
  \city{}
  \country{Singapore}}
\email{cwngo@smu.edu.sg}

\author{Yu-Gang Jiang}
\orcid{0000-0002-1907-8567}
\affiliation{%
  \institution{Shanghai Key Lab of Intell. Info. Processing, School of CS, Fudan University, China}
  \city{}
  \country{}}
\email{ygj@fudan.edu.cn}



\renewcommand{\shortauthors}{Yinxuan Gui, Bin Zhu, Jingjing Chen, Chong-Wah Ngo, \& Yu-Gang Jiang}

\begin{abstract}
Current research in food analysis primarily concentrates on tasks such as food recognition, recipe retrieval and nutrition estimation from a single image. Nevertheless, there is a significant gap in exploring the impact of food intake on physiological indicators (e.g., weight) over time. This paper addresses this gap by introducing the DietDiary dataset, which encompasses daily dietary diaries and corresponding weight measurements of real users. Furthermore, we propose a novel task of weight prediction with a dietary diary that aims to leverage historical food intake and weight to predict future weights. To tackle this task, we propose a model-agnostic time series forecasting framework. Specifically,
we introduce a Unified Meal Representation Learning (UMRL) module to extract representations for each meal. Additionally, we design a diet-aware loss function to associate food intake with weight variations. By conducting experiments on the DietDiary dataset with two state-of-the-art time series forecasting models, NLinear and iTransformer, we demonstrate that our proposed framework achieves superior performance compared to the original models.  We make our dataset, code, and models publicly available at: \href{https://yxg1005.github.io/weight-prediction/}{\textcolor[RGB]{238,21,167}{https://yxg1005.github.io/weight-prediction}}. 
\end{abstract}





\maketitle

\section{Introduction}
Food plays a vital role in human existence, affecting the quality of life and various physiological indicators, with weight being one of the most fundamental and critical aspects. \cite{jequier1999weight-energy-balance} indicates that the relative balance between energy expenditure and dietary intake determines weight gain or loss. Notably, the energy expenditure in non-exercise activities (such as sitting) usually accounts for a significantly larger portion of total energy expenditure than exercise \cite{hamilton2007role-nonexercise}. Motivated by these insights, we aim to investigate the impact of food intake on weight prediction, which can offer valuable insights for individuals aiming to monitor their diet and manage their weight and health effectively over the long term.

Numerous studies have focused on various tasks within the food domain \cite{min2019survey}, such as food classification \cite{chen2017food-classification-1, martinel2018classification-2, jiang2019classification-5, min2023food2k, liu2024canteen}, ingredients recognition \cite{bolanos2017foodingredients-1, chen2016food172, chen2020zeroingredients-2}, recipe retrieval \cite{wang2019retrieval-1, guerrero2021crossretrieval-2, salvador2021revampingretrieval-3, zhu2020cross, shukor2022transformerretrieval-4, wang2021crossretrieval-5, zhu2019r2ganretrieval-6}, food volume prediction \cite{suzuki2020pointvolume-1, lo2019point2volume-2, makhsous2019novelvolume-3} as well as calorie and nutritional information estimation \cite{meyers2015im2calories, thames2021nutrition5k, tai2023nutritionverse}. Nevertheless, these existing methods typically analyze each food image independently and do not investigate the impact of food intake on physiological indicators (e.g., weight) over a period of time.

\begin{figure}[t]
  \includegraphics[width=0.47\textwidth]{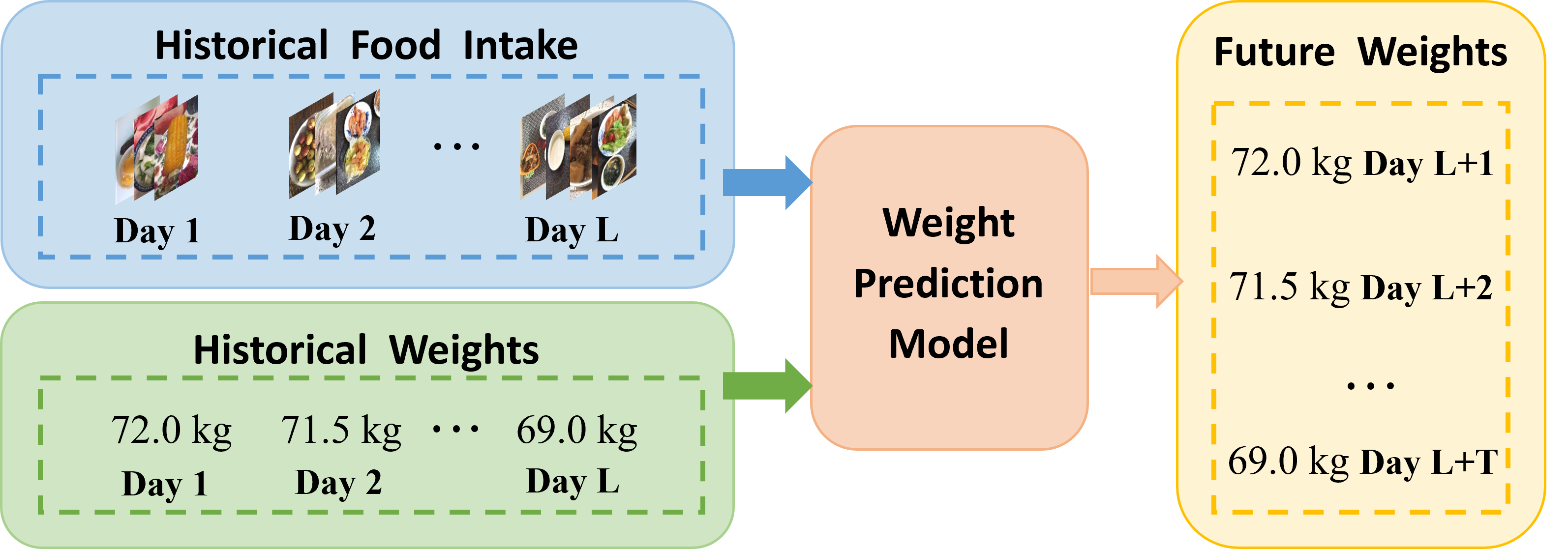}
  \caption{The overview of the proposed weight prediction with diet diary task. Given a historical food intake and corresponding weight measurement in the past $L$ days, the task aims to predict the weights in the following $T$ days.}
  \Description{overview}
  \label{fig:task-overview}
\end{figure}

To address this gap, this paper navigates weight prediction with dietary diary. On the one hand, we construct a novel dataset named DietDiary, which includes daily dietary records and corresponding weight measurements for 611 participants from a health management system.  Different from existing datasets in the food domain such as \cite{bossard2014food101, chen2016food172, min2023food2k}, DietDiary is the first dataset to provide both food intake and corresponding weight measurements over a period of time. On the other hand, as illustrated in Figure \ref{fig:task-overview}, we define a new task that aims to leverage historical weight and food intake data to predict future weights.
This task poses two major challenges: (1) Meal representation learning. Given that food intake can be represented in various forms, such as textual ingredient labels or food images, developing an effective method for learning meal representations for weight prediction is crucial. (2) Modeling the correlation between food intake and weight changes. Since food intake is closely related to weight, weight prediction requires more than just exploring the temporal relationships between historical and future weights. Understanding the complex correlation and dependencies between dietary intake and weight fluctuations presents significant challenges to this task.

To address the aforementioned challenges, we propose a novel framework that integrates food intake information for weight prediction. Specifically, we introduce a Unified Meal Representation Learning (UMRL) module that leverages CLIP~\cite{radford2021learningCLIP} text or image encoders to extract a unified feature representation of a meal from various forms of historical food intake. Additionally, we propose a diet-aware loss function to enable the model to capture the correlation and dependencies between food and weight changes. Importantly, our framework is designed to be compatible with any existing time series forecasting model for weight prediction that incorporates food intake information. We evaluate our framework on the DietDiary dataset using two representative advanced time series forecasting models, NLinear \cite{zeng2023NLinear} and iTransformer \cite{liu2023itransformer}. The superior performance of our framework over these models demonstrates its effectiveness in leveraging food intake information for weight prediction.

The contributions of this paper can be summarized as follows:
\begin{itemize}
  \item We construct a new DietDiary dataset, which is the first dataset providing food intake and corresponding weight measurements over a period of time. 
  \item We introduce a novel task of weight prediction using dietary information. This task uniquely treats food intake as temporal data and investigates its potential to improve weight prediction.
  \item We propose a model-agnostic time series forecasting framework for weight prediction with food intake.  Our framework demonstrates significant improvements over existing methods in weight prediction performance.
\end{itemize}

\section{Related Work}
\subsection{Time Series Forecasting}
Time series forecasting is a fundamental task in various fields, including finance, traffic and meteorology. Over the years, to address the inherent challenges of predicting future values based on past observations, numerous models are proposed, such as the famous ARIMA\cite{arima}. With the rise of deep learning, neural network-based approaches have gained popularity in time series forecasting. Models such as Long Short-Term Memory (LSTM) networks\cite{LSTM1997}, DeepAR\cite{salinas2020deepar} and Prophet\cite{taylor2018Prophet} are used to capture long-term dependencies and nonlinear relationships in the data. As Transformer\cite{liu2023itransformer} demonstrates powerful sequence modeling capabilities in natural language
processing\cite{devlin2018bert}, computer vision\cite{carion2020transformer-in-cv1, vit2020transformer-in-cv2} and other domains, models based on modifying the vanilla Transformer, especially the attention mechanism, tailored for time series forecasting tasks have been widely researched, such as Informer\cite{zhou2021informer}, Autoformer \cite{wu2021autoformer}, Pyraformer \cite{liu2021pyraformer}, and FEDformer \cite{zhou2022fedformer}.

While various research efforts are ongoing to modify attention structures to achieve better Transformer-based solutions, \cite{zeng2023NLinear} questions the effectiveness of the Transformers and proposes a set of simple and efficient linear models including Linear, NLinear, and DLinear, which have led more researchers to focus on linear-based model. TiDE \cite{das2023longTiDE} designs a Multi-layer Perceptron (MLP) based encoder-decoder model to capture covariates and non-linear dependencies. TSMixer \cite{ekambaram2023tsmixer-IBM} patches the time series data and enhances the learning capability of simple MLP structures. 
PatchTST \cite{Yuqietal-2023-PatchTST} uses patched data as input and adopts a channel-independence design that makes a token only contains information from one channel, instead of changing the structure of the transformer. iTransformer \cite{liu2023itransformer} inverts the duties of the attention mechanism and the feed-forward layer to capture temporal information to achieve promoted performance and generalization.

In contrast, we propose a model-agnostic time series forecasting framework to predict weight with food intake. To validate the effectiveness of our framework, state-of-the-art linear-based model NLinear \cite{zeng2023NLinear} and transformer-based model iTransformer \cite{liu2023itransformer} are chosen as our time series forecasting models respectively. 

\begin{figure}[!htbp]
  \includegraphics[width=0.42\textwidth]{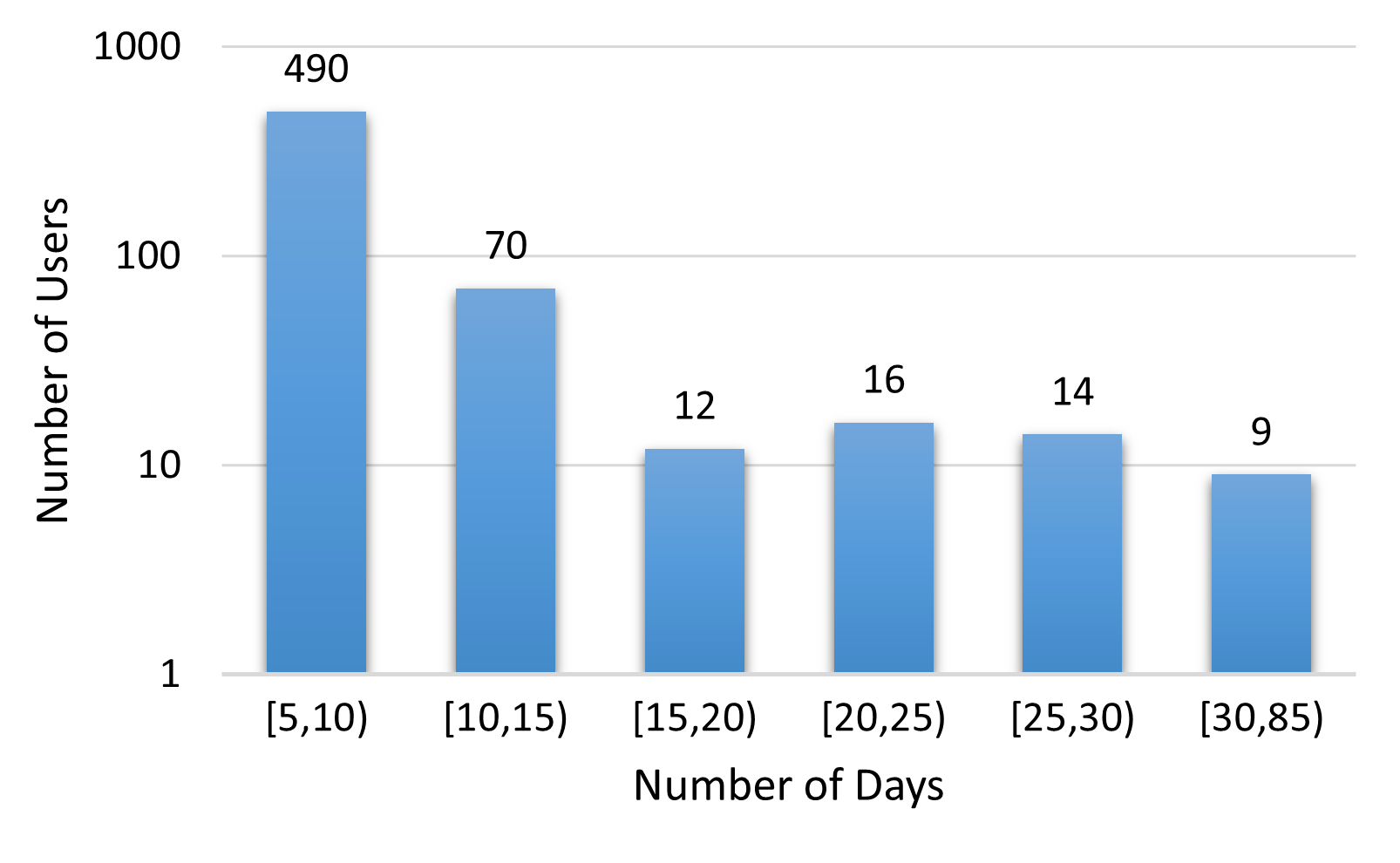}
  \caption{Distribution of the number of recording days for the participants in DietDiary. The y-axis is in the log scale.}
  \Description{Distribution of Users' Recording Days}
  \label{fig:distribution-of-days}
\end{figure}

\begin{figure*}[!htbp]
  \includegraphics[width=\textwidth]{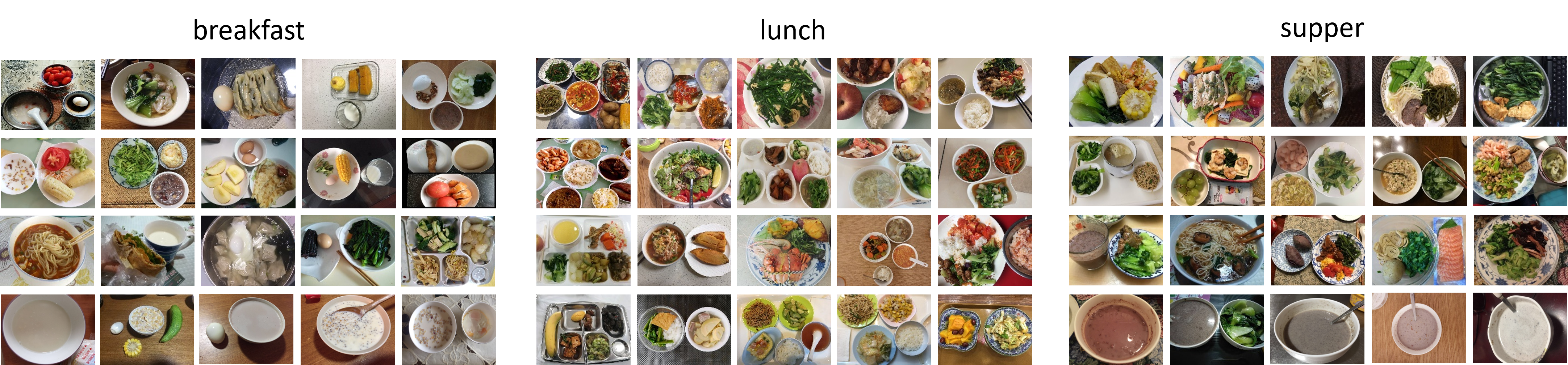}
  \caption{Examples of food images from three meals in DietDiary.}
  \Description{images example}
  \label{fig:image-example}
\end{figure*}

\begin{figure}[!htbp]
  \includegraphics[width=0.475\textwidth]{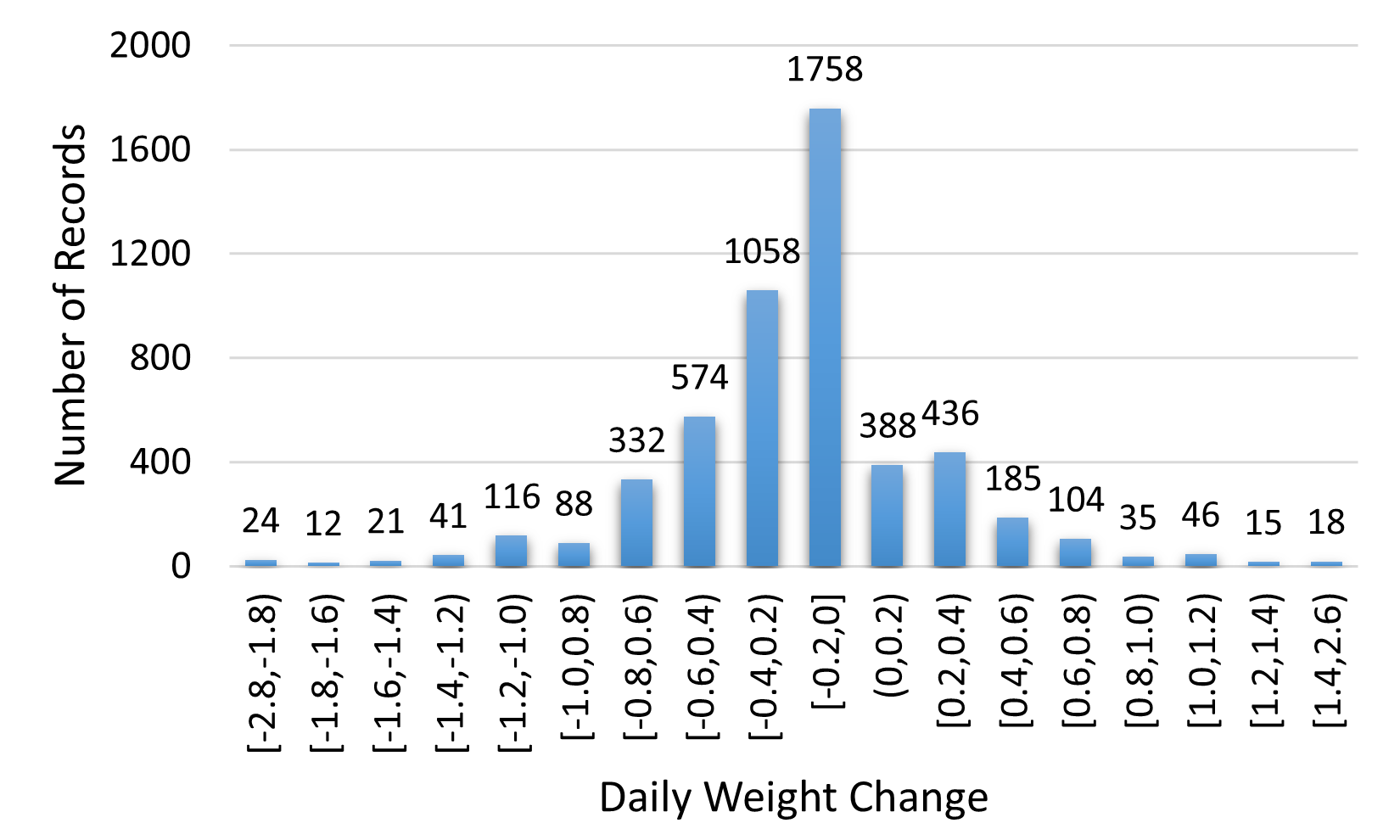}
  \caption{Distribution of Daily Weight Change of all records in DietDiary.}
  \Description{frequency}
  \label{fig:weight-change-single-day}
\end{figure}

\begin{figure}[!htbp]
  \includegraphics[width=0.475\textwidth]{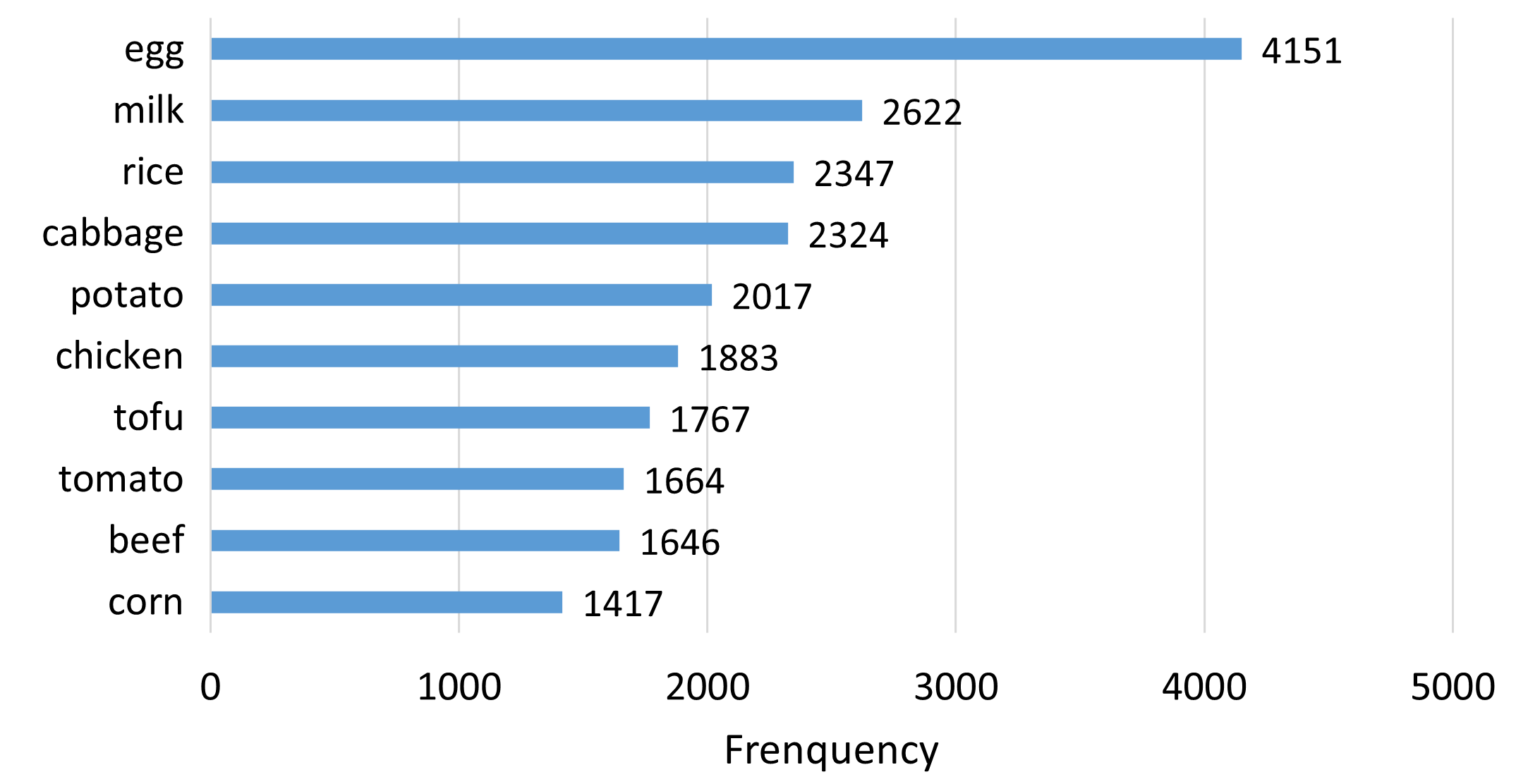}
  \caption{Top 10 ingredients by occurrence frequency in DietDiary.}
  \Description{frequency}
  \label{fig:ingredients-frequency}
\end{figure}

\subsection{Food Analysis}

With the development of computer vision and the emergence of various food datasets, research methods and tasks in the food domain have gradually become more diverse. Some traditional visual tasks have been extended to the food domain, such as food classification \cite{chen2017food-classification-1, martinel2018classification-2, jiang2019classification-5, kiourt2020food-classification-3,  min2023food2k}, ingredients recognition \cite{chen2016food172,bolanos2017foodingredients-1, chen2020zeroingredients-2, liu2020food-ingredients-3}, food detection \cite{zhou2024food-detection-1, aguilar2018food-detection-2}, and food segmentation \cite{freitas2020myfood-seg-1, wu2021large-seg-2, honbu2022unseen-seg-3, lan2023foodsam-seg-4}. With the release of Recipe1M \cite{salvador2017recipe1M}, cross-modal food-related tasks have been extensively studied, such as the retrieval of finding the most relevant recipe for a food image or vice versa \cite{chen2017crossretrieval-7, chen2018deepretrieval-8,zhu2019r2ganretrieval-6,guerrero2021crossretrieval-2, shukor2022transformerretrieval-4, song2024}. Some works focus on generating recipes from images \cite{salvador2019inversecooking, h2020recipegptrecipe-generation-1, chhikara2024fire-recipe-generation-2} or corresponding images from recipes \cite{pan2020chefganfood-img-generation-1, zhu2020cookganfood-img-generation-2}. These tasks all involve analyzing each image independently, whereas in this paper, we model temporal dietary data, which has not been explored in previous studies.

Another research branch involves predicting the nutritional components and calorie of food from food images, which aids in monitoring intake patterns. Some works estimate volume first from voxel \cite{meyers2015im2calories}, point cloud \cite{gao2018foodpoint-cloud, lo2019point2volume-point-cloud-2} or 3D mesh \cite{naritomi2021mesh}, then  mapping is achieved through data on the calories contained per unit volume. With the release of Nutrition5K \cite{thames2021nutrition5k}, a dataset providing fine-grained nutritional attributes, food quality and food videos, some works predict nutrition directly from images by neural network \cite{thames2021nutrition5k, tai2023nutritionverse} or even multimodal large language models \cite{yin2023foodlmm, jiao2024rode}. However, these works only predict nutritional information without further linking food intake to weight. To the best of our knowledge, this paper is the first work to utilize dietary diary to predict weight.

\section{Dataset Construction}

We introduce a novel dataset, DietDiary, specifically for analyzing weight in relation to food intake. In contrast to datasets such as Food-101~\cite{bossard2014food101}, VIREO Food-172~\cite{chen2016food172, chen2020study}, Food2K~\cite{min2023food2k} and Recipe1M~\cite{salvador2017recipe1M}, DietDiary encompasses diet diary of three meals over a period of time, accompanied by daily weight measurement. To the best of our knowledge, DietDiary is the first dataset in the food domain to provide this kind of data, offering new opportunities for research in dietary pattern analysis and its impact on weight management.

\begin{figure*}[htbp]
  \includegraphics[width=0.96\textwidth]{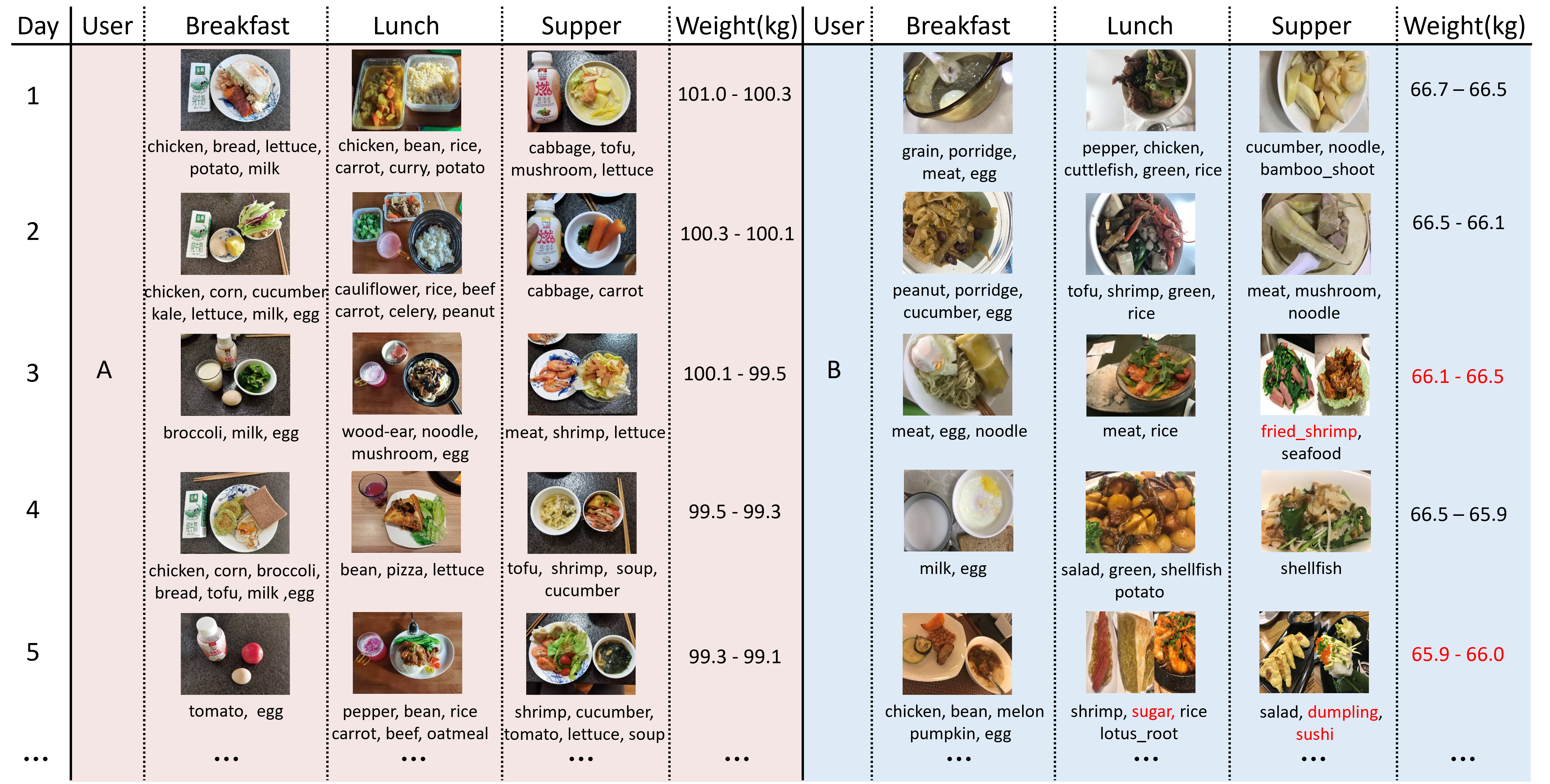}
  \caption{Data records for two participants with different weight fluctuation trends in DietDiary. The records leading to weight gain are highlighted in red.
  }
  \Description{dataset example}
  \label{fig:dataset-example}
\end{figure*}

\subsection{DietDiary Dataset}
The DietDiary dataset was collected within a health management system~\footnote{https://www.qiezilife.com/}
wherein participants were required to meticulously log their daily dietary intake, along with their corresponding weight measurements. The dietary log consists of images of three meals (i.e., breakfast, lunch, and supper) along with manually labeled ingredients for each meal. The dataset encompasses a total of 611 participants, with over 5k daily records, nearly 30k images, and over 15k ingredient annotations. Note that user privacy issues in our dataset have been addressed carefully by anonymizing user information. The
data primarily consists of weights and food images,
without any sensitive personal information.

As depicted in Figure \ref{fig:distribution-of-days}, the duration of record-keeping among participants varies, ranging from a minimum of one week to over one month. A significant majority (91.6\%) of the participants maintained their records for a duration of one to two weeks. However, there were also 9 participants recorded their diet and weight for more than a month. Figure \ref{fig:weight-change-single-day} illustrates the distribution of daily weight fluctuations across all participants. Most users experience weight changes within $\pm 1$ kg per day, with weight loss records comprising the majority due to the data source. Additionally, a small portion of individuals experience changes exceeding 2 kg within a day.

\subsection{Food Images and Ingredient Annotation}
Figure \ref{fig:image-example} presents meal images from various participants, illustrating typical examples of breakfast, lunch, and supper. It is important to note that we have manually curated the dataset to exclude images that are not relevant to food.

The annotation of ingredients in the dataset was standardized through a series of preprocessing steps. First, we unified the separators used between phrases, as different participants employed varying symbols (e.g., ``- ", `` / ", `` () ") to delineate ingredients. Second, the original annotations in Chinese were translated into English. Third, we adopted the approach as in inverse cooking \cite{salvador2019inversecooking} to aggregate and categorize ingredients based on common prefixes or suffixes, for example, "boiled eggs", "steamed eggs", and "egg slices" are all merged into the "egg" category. 
Fourth, ingredients that appeared less than five times were excluded from the dataset. After preprocessing, the total number of unique ingredients was 197. The ten most frequently occurring ingredients are depicted in Figure \ref{fig:ingredients-frequency}, such as "milk", "egg" and "rice", all of which are common in everyday meals. Furthermore, Figure \ref{fig:dataset-example} provides a comprehensive example from the dataset, showcasing food images, annotated ingredients, and corresponding weight measurements for two participants with different weight fluctuation trends. On the one hand, user A maintained a healthy dietary habit, and the weight measurements consistently show a decreasing trend. On the other hand, user B consumed high-calorie foods (such as fried shrimp and sugar) on the third and fifth days, and also consumed a large amount of staple food (dumplings and sushi) during supper, resulting in an increase in corresponding weight records.

\begin{figure*}[htbp]
  \includegraphics[width=1\textwidth]{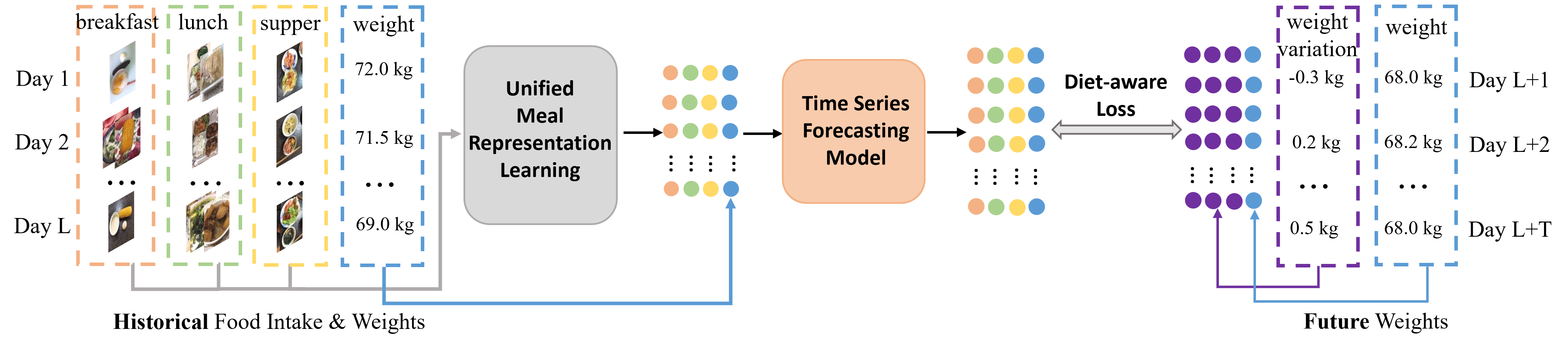}
  \caption{Framework Overview. The ``Unified Meal Representation Learning" module is proposed to map the historical food intake into a time series meal feature sequence. The features and historical weight sequence are combined and subsequently fed into an agnostic time series forecasting model to predict future weights.}

  \Description{overview}
  \label{fig:framework-overview}
\end{figure*}

\begin{figure}[htbp]
  \includegraphics[width=0.48\textwidth]{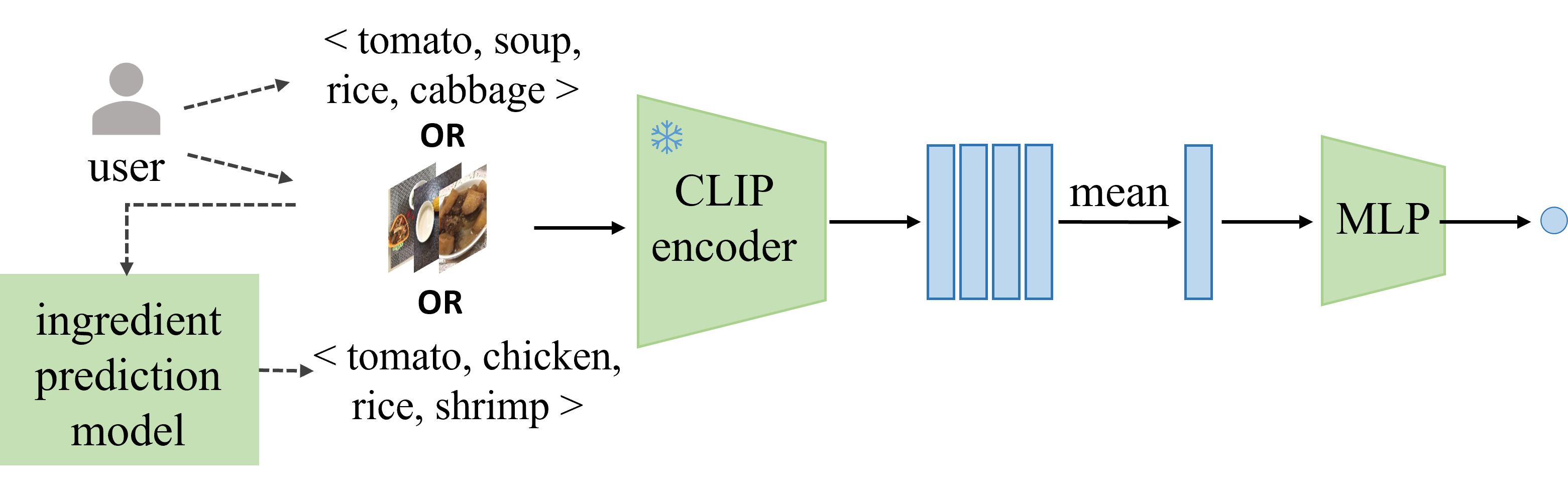}
  \caption{The architecture of proposed Unified Meal Representation Learning Module.}
  \Description{overview}
  \label{fig:UMR}
\end{figure}

\section{Method}

\subsection{Task Definition}
\label{sec:taskDefinition}
Given the historical weight sequence $\textit{W} = \{w_t\}^{L}_{t=1}$ and corresponding food intake $\textit{F} = \{Breakfast_t, Lunch_t, Supper_t\}^{L}_{t=1}$ for the past $L$ days,  the goal of our proposed task is to predict the future weight sequence $\hat{\textit{W}} = \{w_t\}^{L+T}_{t=L+1}$ over the next $T$ days, where $w_t$ denotes the weight value of the t-\textit{th} day, $Breakfast_t$, $Lunch_t$ and $Supper_t$ represent sets of food images or ingredient labels for each meal on the t-\textit{th} day. Note that the historical observation takes various data modalities, including textual ingredient labels or food images, and numerical weight.

\subsection{Proposed Framework}
\label{sec:framework}
Figure \ref{fig:framework-overview} depicts an overview of the proposed framework. The framework primarily consists of three key components: a unified Meal Representation Learning (UMRL) module, an agnostic time series forecasting model, and a diet-aware loss computation.

\textbf{Unified Meal Representation Learning (UMRL)} is a module to extract representation for each meal. Given the historical food intake (e.g., food images or ingredient labels), UMRL is designed to obtain representation for breakfast, lunch, and supper for each day respectively, denoted as $f = \{b_t, l_t, s_t\}^{L}_{t=1}$, where $b_t$, $l_t$ and $s_t$ are representation for breakfast, lunch, and supper on the t-\textit{th} day.
More details of UMRL are presented in Section~\ref{sec:UMRL}.

\textbf{Model-agnostic weight prediction.}
After obtaining the representation of historical food intake from UMRL, we concatenate it with the historical weight $\textit{W}$, denoted as
$\textit{X} = \{b_t, l_t, s_t, w_t\}^{L}_{t=1}$, and feed into a time series forecasting model $M$. The weight prediction result is obtained as follows:
\begin{equation}
  \label{eq:prediction}
  \hat{\textit{Y}} = \{ \hat{b}_t, \hat{l}_t, \hat{s}_t, \hat{w}_t\}^{L+T}_{t=L+1} = \textit{M (X)},
\end{equation}
where $\hat{b}_t$, $\hat{l}_t$, $\hat{s}_t$, $\hat{w}_t$ are the predicted weights from breakfast, lunch, supper and weight respectively. Note that our proposed framework is agnostic to the time series forecasting model, which gives us the flexibility to utilize existing state-of-the-art models for our task. In this paper, we choose linear-based NLinear \cite{zeng2023NLinear} and Transformer-based iTransformer \cite{liu2023itransformer} as $M$ in Equation \ref{eq:prediction}.

\textbf{Diet-aware loss computation.} We introduce a diet-aware loss to model the food intake and weight variations. Assume the weight variation $\Delta_{t}$ of the t-\textit{th} day is calculated by the following formula:
\begin{equation}
  \label{eq:delta}
  \Delta_{t} = w_t - w_{t-1}.
\end{equation}
As each meal contributes to the weight variations, the diet loss is computed by combining all three meals as follows:
\begin{equation}
  \label{eq:l2 loss}
  \mathcal{L}_{diet} = \frac{1}{3} ((\Delta_{t} - \hat{b}_t)^2 + (\Delta_{t} - \hat{l}_t)^2 + (\Delta_{t} - \hat{s}_t)^2).
\end{equation}
In addition, we compute a weight loss based on the weight prediction from historical weight as follows:
\begin{equation}
  \label{eq:l2 loss}
  \mathcal{L}_{weight} = (w_{t} - \hat{w}_t)^2.
\end{equation}

The overall diet-aware loss to train the proposed framework is to combine both diet and weight losses:
\begin{equation}
  \label{eq:food-aware loss}
  \mathcal{L} = \lambda\mathcal{L}_{weight} + (1-\lambda) \mathcal{L}_{diet},
\end{equation}
where $\lambda$ is a hyper-parameter to balance the two losses.

\begin{table}
  \caption{Statistics of training, testing, and validation sets for different settings. L-T refers to the setting of using weight and food intake in $L$ history days to predict the weights of future $T$ days.}
  \label{tab:dataset partition}
  \begin{tabular}{l|c c c c c c}
     \hline
    &\multicolumn{5}{c}{\textbf{settings (L-T)} }\\
    \hline
    \textbf{Data split}&3-3&3-5&3-7/5-5/7-3&5-7&7-7\\
    \hline
    train & 1,535 & 1,010 & 837 & 672 & 514\\
    validation & 221 & 149 & 123 & 74 & 57\\
    test & 440 & 296 & 241 & 216 & 164\\
   \hline
\end{tabular}
\end{table} 

\begin{figure*}[htbp]
  \includegraphics[width=0.96\textwidth]{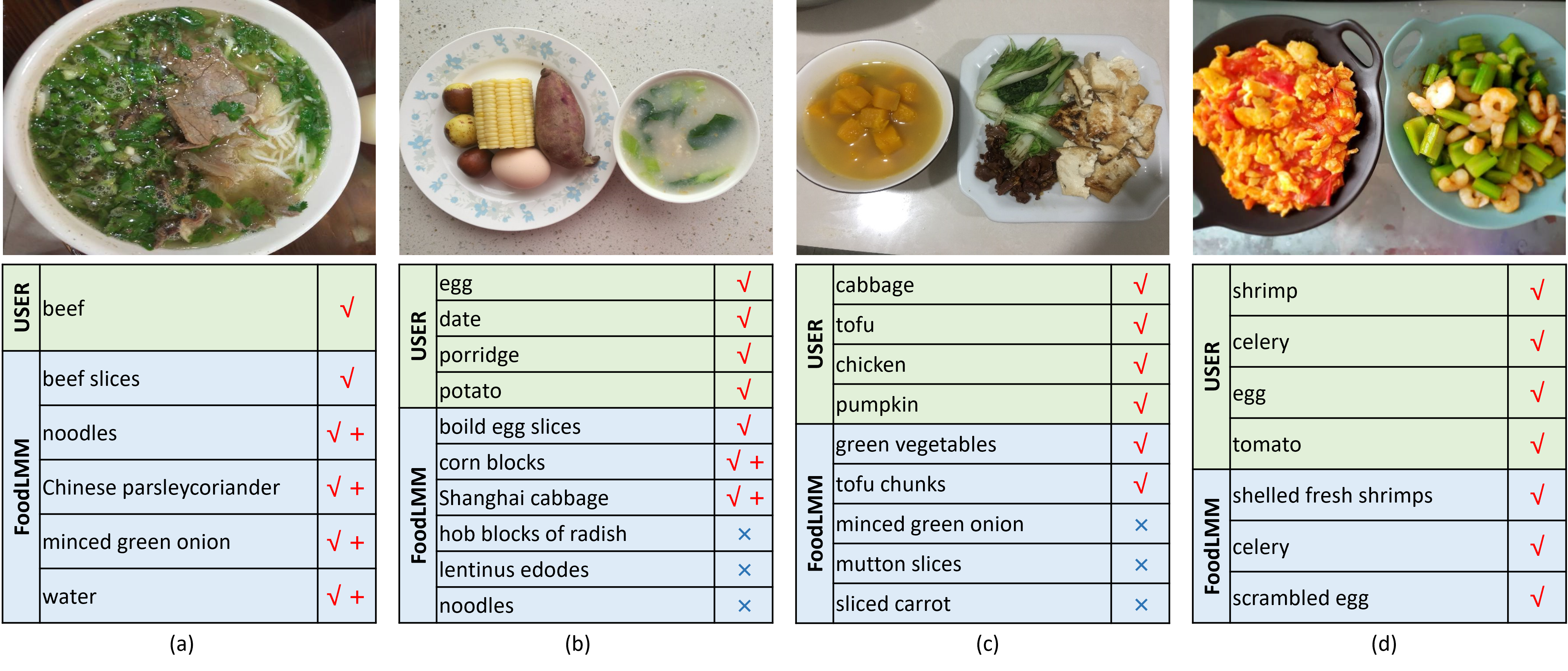}
  \caption{Examples of ingredient labels from user (green) and LMM (blue) for the same food image. The "\textcolor[RGB]{255,0,0}{\checkmark}" and "\textcolor[RGB]{46, 117, 182}{×}" signs indicate the correct and incorrect ingredients respectively. The "\textcolor[RGB]{255,0,0}{\checkmark+}" sign indicates the ingredients from LMM supplement those missed by users.}
  \Description{annotation example}
  \label{fig:User-LMM-ingrs-compare}
\end{figure*}

\subsection{Unified Meal Representation Learning}
\label{sec:UMRL}
As shown in Figure \ref{fig:UMR}, we propose Unified Meal Representation Learning (UMRL) Module to extract representation from historical food intake for each meal. As the food intake could be either food images or ingredient labels, we first employ CLIP~\cite{radford2021learningCLIP} to encode both visual or textual inputs to unify the feature extraction. On the one hand, if the historical food intake is ingredient label, we adopt pretrained CLIP text encoder to extract the features. On the other hand, we adopt CLIP image encoder to extract the features from historical food images. The CLIP model is kept frozen during our training.

Note that the users prefer taking pictures of their meals rather than typing the concrete ingredient annotations in practice. As a result, we also investigate employing a pre-trained ingredient recognition model to automatically obtain ingredient labels from images, which is illustrated in the bottom left of 
The CLIP features of images or ingredient labels are subsequently averaged to get a feature vector for each meal. 
Finally, we feed this vector into a Multilayer Perceptron (MLP) to derive the final meal representation. 
Take breakfast food images as an example,  
denoted $Breakfast_t = \{img_1, ..., img_n\}_{n=1}^{N}$ as a set of breakfast food images on the t-\textit{th} day, where $N$ is the number of images, 
the process of UMRL can be formalized as follows:

\begin{equation}
  \label{eq:UMRL process}
\begin{split}
  E_n = CLIPEncoder(img_n), \\
  \bar{E} = \frac{1}{N} \sum_{n=1}^{N} (E_n), \\
  b_t = MLP(\bar{E}),
\end{split}
\end{equation}
where $E_n$ is the CLIP image feature for $img_n$, $\bar{E}$ indicates the averaged CLIP image features and $b_t$ is the breakfast representation.

\begin{table*}
  \caption{Performance comparison based on NLinear as time series forecasting model. ``image", ``ing-users" and ``ing-LMM" represent the food intake is food image, ingredient labels provided by users, ingredient labels predicted by FoodLMM respectively.}
  \label{tab:experiment result NLinear}
  \resizebox{\textwidth}{!}{\begin{tabular}{c|c  |c c | c c | c c | c c | c c | c c | c c}

     \hline
    \multicolumn{2}{c}{setting} & 
    \multicolumn{2}{|c|}{3 - 3}& \multicolumn{2}{|c|}{3 - 5} & \multicolumn{2}{|c|}{3 - 7}   &  \multicolumn{2}{|c|}{5 - 5} & 
    \multicolumn{2}{|c|}{5 - 7} & 
    \multicolumn{2}{|c|}{7 - 3}  & \multicolumn{2}{|c}{7 - 7} \\
    \hline
    \multicolumn{2}{c|}{metric}  & MSE & MAE& MSE & MAE& MSE & MAE& MSE & MAE& MSE & MAE &MSE &MAE & MSE & MAE\\
    \hline
    \multicolumn{2}{c|}{NLinear~\cite{zeng2023NLinear}} & 
    3.729 & 1.107 &
    5.390 & 1.448 & 
    6.949 & 1.710 & 
    5.991 & 1.619 &
    5.331 & 1.589 &
    5.464 & 1.600 & 
    6.791 & 1.958   \\
    \hline
    \multirow{3}{*}{\rotatebox{90}{\textbf{ours}}} 
    & image & 
    2.769 & 0.985 &
    \textbf{4.765} & \textbf{1.371} & 
    6.332 & 1.636 & 
    2.573 & \textbf{1.330} & 
    \textbf{3.219} & \textbf{1.341} & 
    2.799 & 1.213 & 
    \textbf{2.454} & 1.376  \\
    
    \cline{2-16}
    & ing-users & 
    2.865 & 0.998 &
    4.819 & 1.377 & 
    \textbf{6.307} & \textbf{1.633} & 
    \textbf{2.539} & 1.333 & 
    3.235 & \textbf{1.341} & 
    \textbf{2.742} & \textbf{1.211} & 
    2.479 & 1.370  \\
    \cline{2-16}
    
    & ing-LMM & 
    \textbf{2.048} & \textbf{0.962} &
    4.849 & 1.380 & 
    6.312 & 1.635 & 
    2.542 & 1.345 &
    3.247 & 1.342 & 
    2.968 & 1.233 & 
    2.497 & \textbf{1.365} \\

    \hline

\end{tabular}}
\end{table*}

\begin{table*}
  \caption{Performance comparison based on iTransformer. ``image", ``ing-users" and ``ing-LMM" represent the food intake is food image, ingredient labels provided by users, ingredient labels predicted by FoodLMM respectively.}
  \label{tab:experiment result iTransformer}
  \resizebox{\textwidth}{!}{\begin{tabular}{c|c  |c c | c c | c c | c c | c c | c c | c c}

    \hline
    \multicolumn{2}{c}{setting} & 
    \multicolumn{2}{|c|}{3 - 3}& \multicolumn{2}{|c|}{3 - 5} & \multicolumn{2}{|c|}{3 - 7}   &  \multicolumn{2}{|c|}{5 - 5} & 
    \multicolumn{2}{|c|}{5 - 7} & 
    \multicolumn{2}{|c|}{7 - 3}  & \multicolumn{2}{|c}{7 - 7} \\
    \hline
    \multicolumn{2}{c|}{metric}  & MSE & MAE& MSE & MAE& MSE & MAE& MSE & MAE& MSE & MAE &MSE &MAE & MSE & MAE\\
    \hline
    \multicolumn{2}{c|}{iTransformer~\cite{liu2023itransformer}} & 4.023 & 1.402 &   5.306 & 1.717
  & 5.918 & 1.866   & 5.711 & 1.817 & 6.369 & 1.966 & 4.657 & 1.611 & 4.851 & 1.701 \\
    \hline
    \multirow{3}{*}{\rotatebox{90}{\textbf{ours}}} 
    & image & 
    \textbf{3.436} & \textbf{1.268} &
    5.268 & 1.710 & 
    5.791 & 1.835 & 
    5.478 & 1.783 & 
    4.299 & 1.616 & 
    4.524 & 1.584 & 
    \textbf{4.411} & \textbf{1.596} \\
    
    \cline{2-16}
    & ing-users & 
    3.544 & 1.293 &
    5.202 & 1.696 & 
    5.922 & 1.861 & 
    \textbf{3.540} & \textbf{1.466} & 
    3.746 & 1.541 & 
    4.322 & 1.545 & 
    4.645 & 1.641 \\
    \cline{2-16}
    
    & ing-LMM & 
    4.133 & 1.424 &
    \textbf{5.047} & \textbf{1.662} & 
    \textbf{5.637} & \textbf{1.802} & 
    4.626 & 1.611 & 
    \textbf{3.657} & \textbf{1.532} & 
    \textbf{3.992} & \textbf{1.486} & 
    4.452 & 1.610\\

    \hline

\end{tabular}}
\end{table*}

\section{Experiment}
\subsection{Experiment Settings}
\textbf{Dataset.} 
We conduct experiments on our DietDiary dataset.
Inspired by time series forecasting task~\cite{zeng2023NLinear, liu2023itransformer}, we extensively explore different combinations of using weight and food intake history in $L$ days to predict the weights in future $T$ days, denoted as setting L-T. The settings examined include \{3-3, 3-5, 3-7, 5-5, 5-7, 7-3 and 7-7\}.   For instance, in the 7-3 setting ($L$ = 7, $T$ = 3), we utilize the historical weight sequence and corresponding food intake from the past 7 days to predict the weight for the subsequent 3 days during training.
The dataset is partitioned into training, validation, and test sets in a 7:1:2 ratio for each setting. It is crucial to ensure that each participant appears exclusively in one of these splits, thereby preventing any overlap of participants across training, validation, and testing sets. The settings determine the minimum number of days required for training. To maximize the utility of the dataset, we employ different data partitions for different settings. For example, the settings 5-5 and 7-3 necessitate a minimum of 10 days of dietary intake records per participant, whereas the 7-7 setting requires at least 14 days. Table~\ref{tab:dataset partition} presents the dataset statistics for various settings.

\textbf{Evaluation metrics.}
We employ the Mean Squared Error (MSE) and Mean Absolute Error (MAE) as metrics for performance evaluation. Given that users are primarily concerned with long-term trends in weight prediction, we evaluate performance through autoregressive weight prediction. Specifically, during testing, given a history of $L$ days, we initially forecast the weights for the subsequent $T$ days. Subsequent predictions are then made autoregressively, using the outcomes of previous forecasts as input, until predictions have been generated for all recorded days of users in the test set.

\textbf{Implementation details.}
Given the remarkable performance of 
FoodLMM \cite{yin2023foodlmm}, a Large Multi-modal Model tailored in the food domain, we choose FoodLMM as our ingredient prediction model. 
We evaluate our framework on both the NLinear \cite{zeng2023NLinear} model, representing linear time series forecasting models, and the iTransformer \cite{liu2023itransformer} model, representing transformer-based solutions. 
Both NLinear and iTransformer are trained using Adam optimizer \cite{kingma2014adam} with batch size of 32 until early-stopping criteria is met (validation loss does not decrease after 7 epochs). The learning rate is initially set to 0.005 and decays with each epoch. The $\lambda$ in equation \ref{eq:food-aware loss} is set to \{0, 0.1, 0.25, 0.5, 0.75, 1\} to explore the influence of the proportion of diet loss, which will be shown in Section 5.4.

\subsection{Performance Comparison}

Tables~\ref{tab:experiment result NLinear} and~\ref{tab:experiment result iTransformer} present the performance of our proposed method, which leverages state-of-the-art time series forecasting models NLinear~\cite{zeng2023NLinear} and iTransformer~\cite{liu2023itransformer}, respectively. Compared to models that do not incorporate food intake information, our method consistently achieves superior performance over NLinear~\cite{zeng2023NLinear} and iTransformer~\cite{liu2023itransformer} across all evaluated settings, with a significant margin of improvement. Notably, the improvements are observed consistently with three different types of food intake inputs: food images, user-provided ingredient labels, and ingredient labels predicted by FoodLMM. These results clearly demonstrate the effectiveness of incorporating food intake information in enhancing the accuracy of weight prediction.

An inspiring observation is that the ingredient labels predicted by FoodLMM often perform comparably or even surpass manually provided ingredient labels by users in most settings. This finding is noteworthy considering that acquiring food images is generally easier than manually annotating ingredients for each meal. 
Figure \ref{fig:User-LMM-ingrs-compare} presents four examples comparing ingredient labels between users and FoodLMM. Typically,  users are more prone to omitting ingredients rather than mislabeling them. For instance, "noodles", "Chinese parsleycoriander
", and "green onion" in Figure \ref{fig:User-LMM-ingrs-compare} (a), and ``corn" and ``Shanghai cabbage" in Figure \ref{fig:User-LMM-ingrs-compare} (b) are missing from the user-provided labels. Remarkably, FoodLMM successfully identifies these omitted ingredients. However, ingredient recognition remains a challenging task~\cite{chen2020study}, and the predictions from FoodLMM are still not perfect. For example, there are three incorrect ingredients predicted by FoodLMM in Figure \ref{fig:User-LMM-ingrs-compare} (b) and (c). In Figure \ref{fig:User-LMM-ingrs-compare} (d), although FoodLMM fails to accurately predict ``tomato", the remaining ingredients are correctly identified.

\subsection{Multi-modal Fusion of Food Intake}
We further investigate multi-modal fusions of food intake from both images and ingredient labels by averaging the UMRs of the same meal across different modalities. 
Tables \ref{tab:nlinear-image-ing-users-fusion} and \ref{tab:nlinear-image-ing-lmm-fusion} present the fusion results of images with ingredient annotations from users and FoodLMM respectively. Except for the MSE of setting 5-5, the fusion results consistently outperform those of single modality, whether the ingredient labels are manually provided or predicted by FoodLMM. This demonstrates that food information from different modalities can complement each other, which leads to incorporating both of them achieving better performance. Additionally, the effectiveness of multi-modal fusion also validates that our proposed UMRL module can embed food images and ingredient annotations into a common space and learn unified representations of meals from different modalities.

\subsection{Ablation study}
\textbf{Impact of number of meals.} 
Table \ref{tab:meal-ablation} presents the impact of incorporating different numbers of daily meals on weight prediction performance. Generally, the inclusion of food intake information, regardless of the number of meals, manages to improve performance. When only one meal is considered, the inclusion of lunch data yields the most significant improvement in most settings. This is attributed to lunch meals providing the most diverse and distinct food information within the dataset. Breakfast and supper, often featuring less distinctive items like congee-like foods (e.g., meal replacement powders), do not contribute as significantly as lunch to the encoder's effectiveness. For example, the images in the last row of both breakfast and supper in Figure \ref{fig:image-example}. However, the combination of all three meals offers complementary benefits, leading to the best overall results for weight prediction. 

\begin{table}
  \caption{The multi-modal fusion results of images and ingredient annotations from users based on NLinear.}
  \label{tab:nlinear-image-ing-users-fusion}
  \resizebox{0.475\textwidth}{!}{
  \begin{tabular}{c|c  |c c | c c | c c |cc}

     \hline
    \multicolumn{2}{c}{setting} & 
    \multicolumn{2}{|c|}{5 - 5} & 
    \multicolumn{2}{|c|}{5 - 7} & 
    \multicolumn{2}{|c|}{7 - 3}  & \multicolumn{2}{|c}{7 - 7} \\
    \hline
    \multicolumn{2}{c|}{metric}  & MSE & MAE& MSE & MAE& MSE & MAE& MSE & MAE\\
    \hline
    \multicolumn{2}{c|}{NLinear~\cite{zeng2023NLinear}} & 
    5.991 & 1.619 &
    5.331 & 1.589 & 
    5.464 & 1.600 & 
    6.791 & 1.958 \\
    \hline
    \multirow{3}{*}{\rotatebox{90}{\textbf{ours}}} 
    & image & 
    2.573 & 1.330&
    2.751 & 1.469 &
    3.722 & 1.340 & 
    4.390 & 1.576  \\
    
    \cline{2-10}
    & ing-users & 
    \textbf{2.539} & 1.333 &
    2.773 & 1.480 &
    4.327 & 1.430 &
    4.340 & 1.567 \\
    \cline{2-10}
    
    & fusion & 
    2.567 & \textbf{1.319} &
    \textbf{2.748} & \textbf{1.464} &
    \textbf{3.623} & \textbf{1.323} &
    \textbf{4.334} & \textbf{1.566}
    \\

    \hline

\end{tabular}
}
\end{table}

\begin{table}
  \caption{The multi-modal fusion results of images and ingredient annotations from FoodLMM based on NLinear.}
  \label{tab:nlinear-image-ing-lmm-fusion}
  \resizebox{0.45\textwidth}{!}{
  \begin{tabular}{c|c  |c c | c c | c c |cc}

     \hline
    \multicolumn{2}{c}{setting} & 
    \multicolumn{2}{|c|}{5 - 5} & 
    \multicolumn{2}{|c|}{5 - 7} & 
    \multicolumn{2}{|c|}{7 - 3}  & \multicolumn{2}{|c}{7 - 7} \\
    \hline
    \multicolumn{2}{c|}{metric}  & MSE & MAE& MSE & MAE& MSE & MAE& MSE & MAE\\
    \hline
    \multicolumn{2}{c|}{NLinear~\cite{zeng2023NLinear}} & 
    5.991 & 1.619 &
    5.331 & 1.589 & 
    5.464 & 1.600 & 
    6.791 & 1.958 \\
    \hline
    \multirow{3}{*}{\rotatebox{90}{\textbf{ours}}} 
    & image & 
    2.573 & 1.330&
    2.751 & 1.469 &
    3.722 & 1.340 & 
    4.390 & 1.576  \\
    
    \cline{2-10}
    & ing-LMM & 
    \textbf{2.542} & 1.345 &
    2.765 & 1.476 &
    4.523 & 1.461 &
    4.367 & 1.572 \\
    \cline{2-10}
    
    & fusion & 
    2.563 & \textbf{1.325} &
    \textbf{2.741} & \textbf{1.460} &
    \textbf{3.653} & \textbf{1.328} &
    \textbf{4.361} & \textbf{1.571}
    \\

    \hline

\end{tabular}
}
\end{table} 

\begin{table}
  \caption{Ablation study of number of meals as food intake based on NLinear as time series forecasting model. Food images are used as food intake and MAE is reported.}
   \resizebox{0.45\textwidth}{!}{
  \label{tab:meal-ablation}
  \begin{tabular}{l|ccccc}
    \hline
    setting & 3-3&3-5& 3-7 & 5-7 \\
    \hline
    NLinear~\cite{zeng2023NLinear} & 1.107  
    &1.448
    & 1.710 
    & 1.589 \\
    \hline
    \hspace{3.3em} + breakfast 
    & 0.990 
    &1.407
    &  1.701
    & 1.452 \\
    \hspace{3.3em} + lunch 
    & 0.986 
    & 1.408
    & 1.692 
    & 1.450 \\
    \hspace{3.3em} + supper 
    & 1.032 
    & 1.413
    & 1.697 
    & 1.450 \\
    \hline
    \hspace{3.3em} + breakfast + lunch 
    & 1.004 
    &1.404
    & 1.695 
    & 1.375 \\
    \hspace{3.3em} + breakfast + supper 
    & 0.987 
    & 1.424
    & 1.702 
    & 1.373 \\
    \hspace{3.3em} + lunch + supper 
    & 1.012 
    & 1.410
    & 1.686 
    & 1.374\\
    \hline
   \hspace{3.3em} + all three meals & \textbf{0.985} & \textbf{1.371} & \textbf{1.636} & \textbf{1.341}\\
    \hline
  \end{tabular}
  }
\end{table}

\begin{figure}[htbp]
  \includegraphics[width=0.48\textwidth]{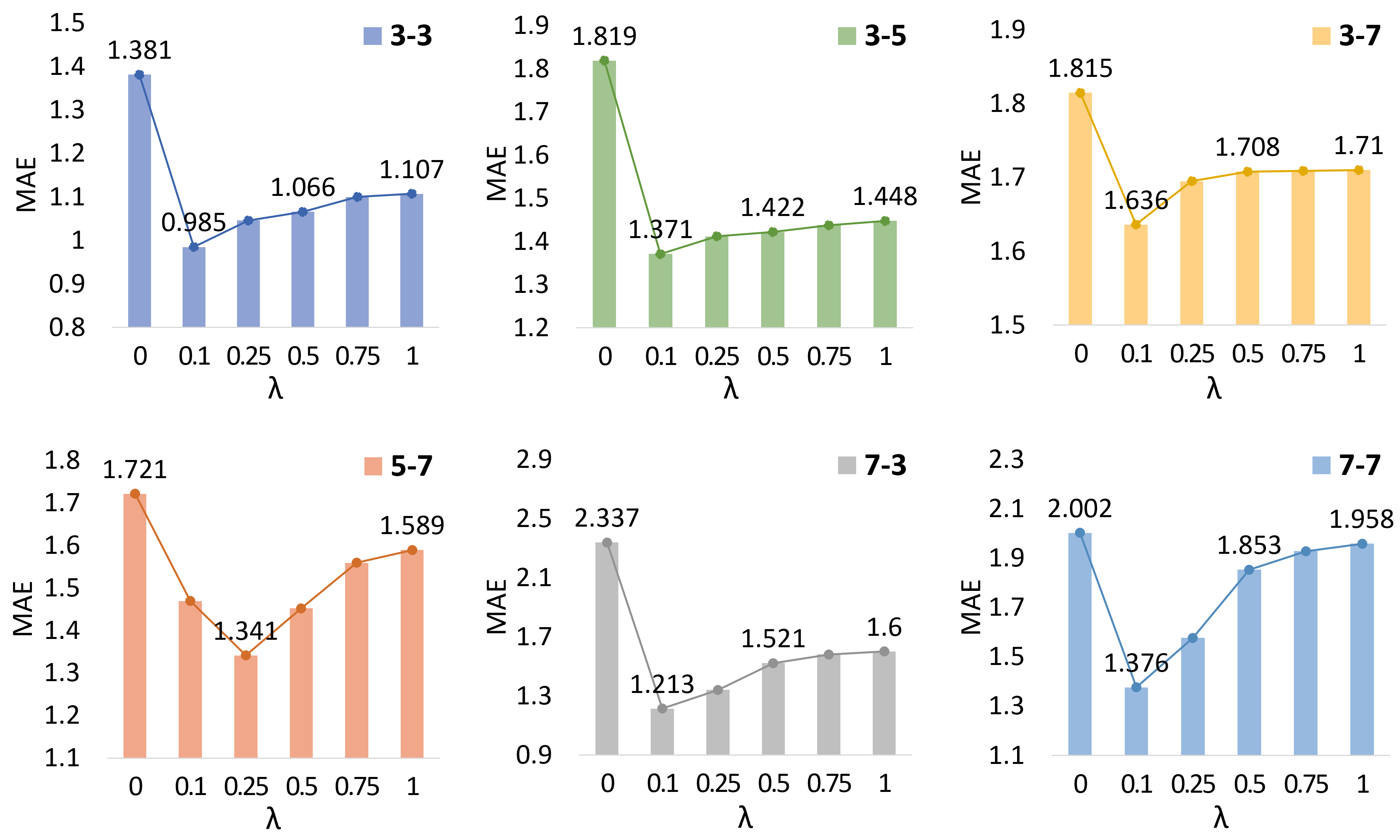}
  \caption{Ablation study of hyper-parameter of $\lambda$ based on NLinear model in terms of MAE.}
  \Description{frequency}
  \label{fig:lambda-ablation}
\end{figure}

\noindent\textbf{Hyper-parameter $\lambda$.}
We further conduct an ablation study to investigate the impact of the hyper-parameter $\lambda$ in Equation \ref{eq:food-aware loss}, which balances the contributions of the weight loss $\mathcal{L}_{weight}$ and the diet loss $\mathcal{L}_{diet}$. Figure \ref{fig:lambda-ablation} illustrates the performance of weight prediction under different settings with $\lambda$ values ranging from 0 to 1, specifically $\lambda \in$ \{0, 0.1, 0.25, 0.5, 0.75, 1\}.
As $\lambda$ increases from 0 to 1, which corresponds to a decreasing emphasis on $\mathcal{L}_{diet}$, we observe an initial improvement in performance followed by a gradual decline. The best results are achieved at $\lambda = 0.1$ in most of the settings, demonstrating the robustness of the hyper-parameters across settings. Notably, when $\lambda = 1$, only $\mathcal{L}_{weight}$ is used to optimize the model, which is equivalent to NLinear model by excluding food intake. On the contrary, when $\lambda = 0$ and the model is solely optimized using $\mathcal{L}_{diet}$, resulting in inferior performance by combining the weight loss. The results show the significance of integrating both weight and diet losses for effective weight prediction.

\subsection{Weight Prediction Visualization}
\begin{figure}[htbp]
  \includegraphics[width=0.45\textwidth]{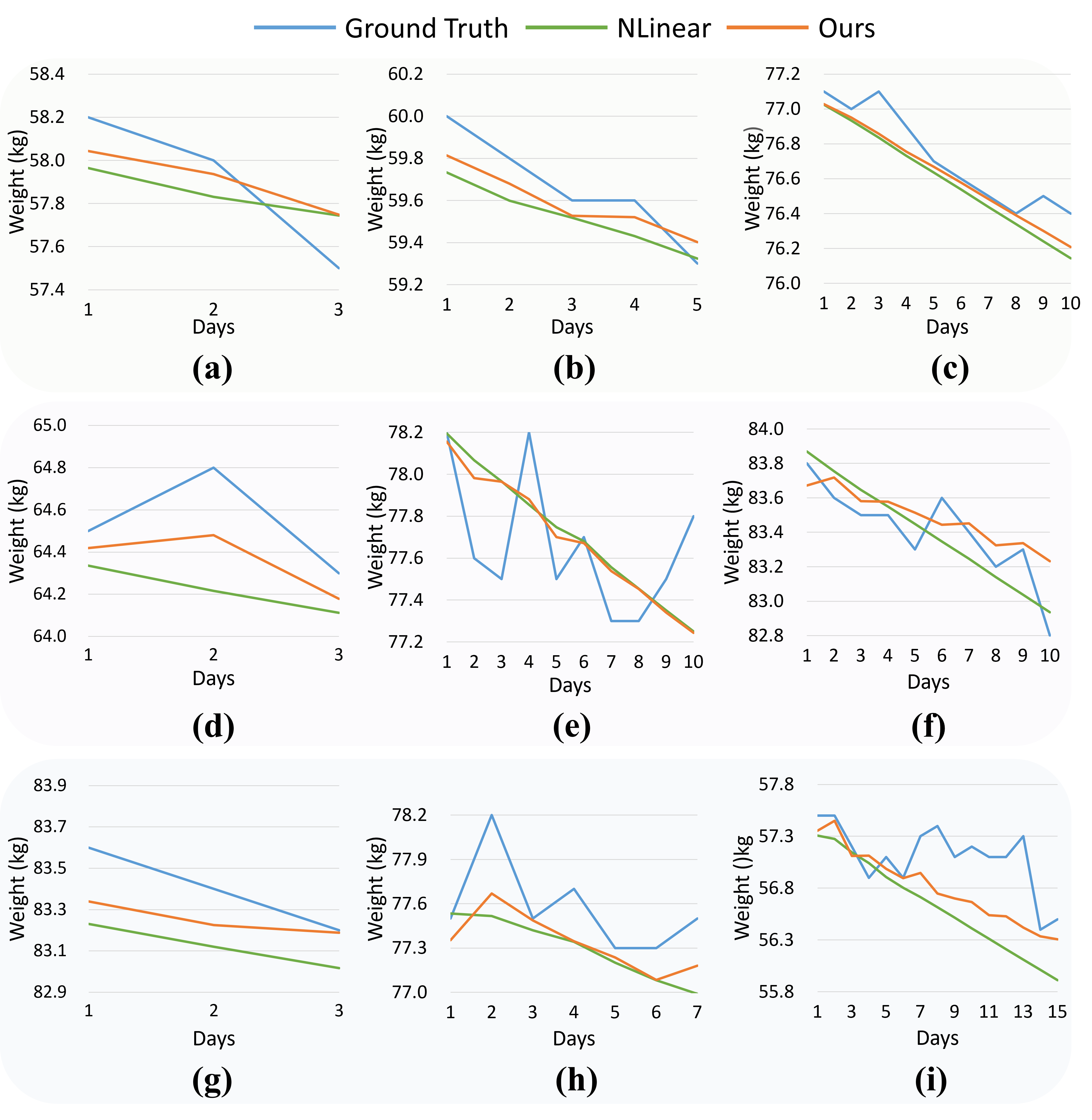}
  \caption{Weight prediction visualization. In each subplot, the x-axis represents the number of predicted days, and the y-axis refers to the weight(kg). The settings for 3-3, 5-5, and 7-3 are displayed in the first (i.e., (a), (b) and (c)), second (i.e., (d), (e) and (f)), and third (i.e., (g), (h) and (i)) rows respectively.}
  \Description{frequency}
  \label{fig:visualization}
\end{figure}

As illustrated in Figure \ref{fig:visualization}, we qualitatively compare the weight prediction visualization among ground-truth weight (blue), NLinear model (green), and our framework based on NLinear (orange) using images as dietary information for different users. It is evident that both the trend and the exact predicted values of our framework are closer to the ground truth than those of the NLinear model. For instance, in subplot (b),
the ground truth weight trend shows an initial decrease, followed by a slight increase, and then another decrease. The orange line, representing our prediction results, exhibits the same trend and is numerically closer to the ground truth than the green line of the NLinear model. Similar trends can be observed in subplots (a), (d), (f), and (g) as well. However, in subplots (c) and (i), even though our prediction results are superior to the NLinear model, our framework's predictions do not align well with the ground truth trend. This discrepancy can be attributed to variable changes in weight and the accumulation of prediction errors in long-term predictions. This issue remains an unresolved challenge in time series forecasting tasks \cite{zeng2023NLinear, wang2020long-term-error}.

\section{Conclusion}
We have investigated weight prediction using diet diary. A new dataset named DietDiary is constructed which comprises dietary intake and corresponding daily weight measurements over a period. We introduce a new task of predicting weights by leveraging historical food intake. To address this task, we propose a model-agnostic time series forecasting framework that achieves significant improvements in weight prediction. Our experiments conducted on two representative time series prediction models, NLinear \cite{zeng2023NLinear} and iTransformer \cite{liu2023itransformer}, not only highlight the effectiveness of the proposed UMRL module and diet-aware loss, but also demonstrate that incorporating food intake leads to improved accuracy in weight prediction across various forms of intake information, including images, ingredient annotations provided by users, and ingredient labels predicted from a pre-trained ingredient prediction model or the multimodal combination with both image and ingredient.
While encouraging, this paper primarily focuses on weight prediction from food intake, exploring the impact of food to other physiological indicator such as blood glucose, will be our future work.

\begin{acks}
This work was supported by NSFC project (No. 62072116).
\end{acks}

\bibliographystyle{ACM-Reference-Format}

\bibliography{sample-sigconf}


\begin{thebibliography}{67}


\ifx \showCODEN    \undefined \def \showCODEN     #1{\unskip}     \fi
\ifx \showDOI      \undefined \def \showDOI       #1{#1}\fi
\ifx \showISBNx    \undefined \def \showISBNx     #1{\unskip}     \fi
\ifx \showISBNxiii \undefined \def \showISBNxiii  #1{\unskip}     \fi
\ifx \showISSN     \undefined \def \showISSN      #1{\unskip}     \fi
\ifx \showLCCN     \undefined \def \showLCCN      #1{\unskip}     \fi
\ifx \shownote     \undefined \def \shownote      #1{#1}          \fi
\ifx \showarticletitle \undefined \def \showarticletitle #1{#1}   \fi
\ifx \showURL      \undefined \def \showURL       {\relax}        \fi
\providecommand\bibfield[2]{#2}
\providecommand\bibinfo[2]{#2}
\providecommand\natexlab[1]{#1}
\providecommand\showeprint[2][]{arXiv:#2}

\bibitem[Aguilar et~al\mbox{.}(2018)]%
        {aguilar2018food-detection-2}
\bibfield{author}{\bibinfo{person}{Eduardo Aguilar}, \bibinfo{person}{Beatriz Remeseiro}, \bibinfo{person}{Marc Bola{\~n}os}, {and} \bibinfo{person}{Petia Radeva}.} \bibinfo{year}{2018}\natexlab{}.
\newblock \showarticletitle{Grab, pay, and eat: Semantic food detection for smart restaurants}.
\newblock \bibinfo{journal}{\emph{IEEE Transactions on Multimedia}} \bibinfo{volume}{20}, \bibinfo{number}{12} (\bibinfo{year}{2018}), \bibinfo{pages}{3266--3275}.
\newblock


\bibitem[Bola{\~n}os et~al\mbox{.}(2017)]%
        {bolanos2017foodingredients-1}
\bibfield{author}{\bibinfo{person}{Marc Bola{\~n}os}, \bibinfo{person}{Aina Ferr{\`a}}, {and} \bibinfo{person}{Petia Radeva}.} \bibinfo{year}{2017}\natexlab{}.
\newblock \showarticletitle{Food ingredients recognition through multi-label learning}. In \bibinfo{booktitle}{\emph{New Trends in Image Analysis and Processing--ICIAP 2017: ICIAP International Workshops, WBICV, SSPandBE, 3AS, RGBD, NIVAR, IWBAAS, and MADiMa 2017, Catania, Italy, September 11-15, 2017, Revised Selected Papers 19}}. Springer, \bibinfo{pages}{394--402}.
\newblock


\bibitem[Bossard et~al\mbox{.}(2014)]%
        {bossard2014food101}
\bibfield{author}{\bibinfo{person}{Lukas Bossard}, \bibinfo{person}{Matthieu Guillaumin}, {and} \bibinfo{person}{Luc Van~Gool}.} \bibinfo{year}{2014}\natexlab{}.
\newblock \showarticletitle{Food-101--mining discriminative components with random forests}. In \bibinfo{booktitle}{\emph{Computer Vision--ECCV 2014: 13th European Conference, Zurich, Switzerland, September 6-12, 2014, Proceedings, Part VI 13}}. Springer, \bibinfo{pages}{446--461}.
\newblock


\bibitem[Box et~al\mbox{.}(2015)]%
        {arima}
\bibfield{author}{\bibinfo{person}{George~EP Box}, \bibinfo{person}{Gwilym~M Jenkins}, \bibinfo{person}{Gregory~C Reinsel}, {and} \bibinfo{person}{Greta~M Ljung}.} \bibinfo{year}{2015}\natexlab{}.
\newblock \bibinfo{booktitle}{\emph{Time series analysis: forecasting and control}}.
\newblock \bibinfo{publisher}{John Wiley \& Sons}.
\newblock


\bibitem[Carion et~al\mbox{.}(2020)]%
        {carion2020transformer-in-cv1}
\bibfield{author}{\bibinfo{person}{Nicolas Carion}, \bibinfo{person}{Francisco Massa}, \bibinfo{person}{Gabriel Synnaeve}, \bibinfo{person}{Nicolas Usunier}, \bibinfo{person}{Alexander Kirillov}, {and} \bibinfo{person}{Sergey Zagoruyko}.} \bibinfo{year}{2020}\natexlab{}.
\newblock \showarticletitle{End-to-end object detection with transformers}. In \bibinfo{booktitle}{\emph{European conference on computer vision}}. Springer, \bibinfo{pages}{213--229}.
\newblock


\bibitem[Chen and Ngo(2016)]%
        {chen2016food172}
\bibfield{author}{\bibinfo{person}{Jingjing Chen} {and} \bibinfo{person}{Chong-Wah Ngo}.} \bibinfo{year}{2016}\natexlab{}.
\newblock \showarticletitle{Deep-based ingredient recognition for cooking recipe retrieval}. In \bibinfo{booktitle}{\emph{Proceedings of the 24th ACM international conference on Multimedia}}. \bibinfo{pages}{32--41}.
\newblock


\bibitem[Chen et~al\mbox{.}(2020a)]%
        {chen2020zeroingredients-2}
\bibfield{author}{\bibinfo{person}{Jingjing Chen}, \bibinfo{person}{Liangming Pan}, \bibinfo{person}{Zhipeng Wei}, \bibinfo{person}{Xiang Wang}, \bibinfo{person}{Chong-Wah Ngo}, {and} \bibinfo{person}{Tat-Seng Chua}.} \bibinfo{year}{2020}\natexlab{a}.
\newblock \showarticletitle{Zero-shot ingredient recognition by multi-relational graph convolutional network}. In \bibinfo{booktitle}{\emph{Proceedings of the AAAI Conference on Artificial Intelligence}}, Vol.~\bibinfo{volume}{34}. \bibinfo{pages}{10542--10550}.
\newblock


\bibitem[Chen et~al\mbox{.}(2020b)]%
        {chen2020study}
\bibfield{author}{\bibinfo{person}{Jingjing Chen}, \bibinfo{person}{Bin Zhu}, \bibinfo{person}{Chong-Wah Ngo}, \bibinfo{person}{Tat-Seng Chua}, {and} \bibinfo{person}{Yu-Gang Jiang}.} \bibinfo{year}{2020}\natexlab{b}.
\newblock \showarticletitle{A study of multi-task and region-wise deep learning for food ingredient recognition}.
\newblock \bibinfo{journal}{\emph{IEEE Transactions on Image Processing}}  \bibinfo{volume}{30} (\bibinfo{year}{2020}), \bibinfo{pages}{1514--1526}.
\newblock


\bibitem[Chen et~al\mbox{.}(2017a)]%
        {chen2017crossretrieval-7}
\bibfield{author}{\bibinfo{person}{Jing-jing Chen}, \bibinfo{person}{Chong-Wah Ngo}, {and} \bibinfo{person}{Tat-Seng Chua}.} \bibinfo{year}{2017}\natexlab{a}.
\newblock \showarticletitle{Cross-modal recipe retrieval with rich food attributes}. In \bibinfo{booktitle}{\emph{Proceedings of the 25th ACM international conference on Multimedia}}. \bibinfo{pages}{1771--1779}.
\newblock


\bibitem[Chen et~al\mbox{.}(2018)]%
        {chen2018deepretrieval-8}
\bibfield{author}{\bibinfo{person}{Jing-Jing Chen}, \bibinfo{person}{Chong-Wah Ngo}, \bibinfo{person}{Fu-Li Feng}, {and} \bibinfo{person}{Tat-Seng Chua}.} \bibinfo{year}{2018}\natexlab{}.
\newblock \showarticletitle{Deep understanding of cooking procedure for cross-modal recipe retrieval}. In \bibinfo{booktitle}{\emph{Proceedings of the 26th ACM international conference on Multimedia}}. \bibinfo{pages}{1020--1028}.
\newblock


\bibitem[Chen et~al\mbox{.}(2017b)]%
        {chen2017food-classification-1}
\bibfield{author}{\bibinfo{person}{Xin Chen}, \bibinfo{person}{Yu Zhu}, \bibinfo{person}{Hua Zhou}, \bibinfo{person}{Liang Diao}, {and} \bibinfo{person}{Dongyan Wang}.} \bibinfo{year}{2017}\natexlab{b}.
\newblock \showarticletitle{Chinesefoodnet: A large-scale image dataset for chinese food recognition}.
\newblock \bibinfo{journal}{\emph{arXiv preprint arXiv:1705.02743}} (\bibinfo{year}{2017}).
\newblock


\bibitem[Chhikara et~al\mbox{.}(2024)]%
        {chhikara2024fire-recipe-generation-2}
\bibfield{author}{\bibinfo{person}{Prateek Chhikara}, \bibinfo{person}{Dhiraj Chaurasia}, \bibinfo{person}{Yifan Jiang}, \bibinfo{person}{Omkar Masur}, {and} \bibinfo{person}{Filip Ilievski}.} \bibinfo{year}{2024}\natexlab{}.
\newblock \showarticletitle{Fire: Food image to recipe generation}. In \bibinfo{booktitle}{\emph{Proceedings of the IEEE/CVF Winter Conference on Applications of Computer Vision}}. \bibinfo{pages}{8184--8194}.
\newblock


\bibitem[Das et~al\mbox{.}(2023)]%
        {das2023longTiDE}
\bibfield{author}{\bibinfo{person}{Abhimanyu Das}, \bibinfo{person}{Weihao Kong}, \bibinfo{person}{Andrew Leach}, \bibinfo{person}{Rajat Sen}, {and} \bibinfo{person}{Rose Yu}.} \bibinfo{year}{2023}\natexlab{}.
\newblock \showarticletitle{Long-term forecasting with tide: Time-series dense encoder}.
\newblock \bibinfo{journal}{\emph{arXiv preprint arXiv:2304.08424}} (\bibinfo{year}{2023}).
\newblock


\bibitem[Devlin et~al\mbox{.}(2018)]%
        {devlin2018bert}
\bibfield{author}{\bibinfo{person}{Jacob Devlin}, \bibinfo{person}{Ming-Wei Chang}, \bibinfo{person}{Kenton Lee}, {and} \bibinfo{person}{Kristina Toutanova}.} \bibinfo{year}{2018}\natexlab{}.
\newblock \showarticletitle{Bert: Pre-training of deep bidirectional transformers for language understanding}.
\newblock \bibinfo{journal}{\emph{arXiv preprint arXiv:1810.04805}} (\bibinfo{year}{2018}).
\newblock


\bibitem[Dosovitskiy et~al\mbox{.}(2020)]%
        {vit2020transformer-in-cv2}
\bibfield{author}{\bibinfo{person}{Alexey Dosovitskiy}, \bibinfo{person}{Lucas Beyer}, \bibinfo{person}{Alexander Kolesnikov}, \bibinfo{person}{Dirk Weissenborn}, \bibinfo{person}{Xiaohua Zhai}, \bibinfo{person}{Thomas Unterthiner}, \bibinfo{person}{Mostafa Dehghani}, \bibinfo{person}{Matthias Minderer}, \bibinfo{person}{Georg Heigold}, \bibinfo{person}{Sylvain Gelly}, {et~al\mbox{.}}} \bibinfo{year}{2020}\natexlab{}.
\newblock \showarticletitle{An image is worth 16x16 words: Transformers for image recognition at scale}.
\newblock \bibinfo{journal}{\emph{arXiv preprint arXiv:2010.11929}} (\bibinfo{year}{2020}).
\newblock


\bibitem[Ekambaram et~al\mbox{.}(2023)]%
        {ekambaram2023tsmixer-IBM}
\bibfield{author}{\bibinfo{person}{Vijay Ekambaram}, \bibinfo{person}{Arindam Jati}, \bibinfo{person}{Nam Nguyen}, \bibinfo{person}{Phanwadee Sinthong}, {and} \bibinfo{person}{Jayant Kalagnanam}.} \bibinfo{year}{2023}\natexlab{}.
\newblock \showarticletitle{Tsmixer: Lightweight mlp-mixer model for multivariate time series forecasting}. In \bibinfo{booktitle}{\emph{Proceedings of the 29th ACM SIGKDD Conference on Knowledge Discovery and Data Mining}}. \bibinfo{pages}{459--469}.
\newblock


\bibitem[Freitas et~al\mbox{.}(2020)]%
        {freitas2020myfood-seg-1}
\bibfield{author}{\bibinfo{person}{Charles~NC Freitas}, \bibinfo{person}{Filipe~R Cordeiro}, {and} \bibinfo{person}{Valmir Macario}.} \bibinfo{year}{2020}\natexlab{}.
\newblock \showarticletitle{Myfood: A food segmentation and classification system to aid nutritional monitoring}. In \bibinfo{booktitle}{\emph{2020 33rd SIBGRAPI Conference on Graphics, Patterns and Images (SIBGRAPI)}}. IEEE, \bibinfo{pages}{234--239}.
\newblock


\bibitem[Gao et~al\mbox{.}(2018)]%
        {gao2018foodpoint-cloud}
\bibfield{author}{\bibinfo{person}{Anqi Gao}, \bibinfo{person}{Frank P-W Lo}, {and} \bibinfo{person}{Benny Lo}.} \bibinfo{year}{2018}\natexlab{}.
\newblock \showarticletitle{Food volume estimation for quantifying dietary intake with a wearable camera}. In \bibinfo{booktitle}{\emph{2018 IEEE 15th International Conference on Wearable and Implantable Body Sensor Networks (BSN)}}. IEEE, \bibinfo{pages}{110--113}.
\newblock


\bibitem[Guerrero et~al\mbox{.}(2021)]%
        {guerrero2021crossretrieval-2}
\bibfield{author}{\bibinfo{person}{Ricardo Guerrero}, \bibinfo{person}{Hai~X Pham}, {and} \bibinfo{person}{Vladimir Pavlovic}.} \bibinfo{year}{2021}\natexlab{}.
\newblock \showarticletitle{Cross-modal retrieval and synthesis (x-mrs): Closing the modality gap in shared subspace learning}. In \bibinfo{booktitle}{\emph{Proceedings of the 29th ACM International Conference on Multimedia}}. \bibinfo{pages}{3192--3201}.
\newblock


\bibitem[H.~Lee et~al\mbox{.}(2020)]%
        {h2020recipegptrecipe-generation-1}
\bibfield{author}{\bibinfo{person}{Helena H.~Lee}, \bibinfo{person}{Ke Shu}, \bibinfo{person}{Palakorn Achananuparp}, \bibinfo{person}{Philips~Kokoh Prasetyo}, \bibinfo{person}{Yue Liu}, \bibinfo{person}{Ee-Peng Lim}, {and} \bibinfo{person}{Lav~R Varshney}.} \bibinfo{year}{2020}\natexlab{}.
\newblock \showarticletitle{RecipeGPT: Generative pre-training based cooking recipe generation and evaluation system}. In \bibinfo{booktitle}{\emph{Companion Proceedings of the Web Conference 2020}}. \bibinfo{pages}{181--184}.
\newblock


\bibitem[Hamilton et~al\mbox{.}(2007)]%
        {hamilton2007role-nonexercise}
\bibfield{author}{\bibinfo{person}{Marc~T Hamilton}, \bibinfo{person}{Deborah~G Hamilton}, {and} \bibinfo{person}{Theodore~W Zderic}.} \bibinfo{year}{2007}\natexlab{}.
\newblock \showarticletitle{Role of low energy expenditure and sitting in obesity, metabolic syndrome, type 2 diabetes, and cardiovascular disease}.
\newblock \bibinfo{journal}{\emph{Diabetes}} \bibinfo{volume}{56}, \bibinfo{number}{11} (\bibinfo{year}{2007}), \bibinfo{pages}{2655--2667}.
\newblock


\bibitem[Hochreiter and Schmidhuber(1997)]%
        {LSTM1997}
\bibfield{author}{\bibinfo{person}{Sepp Hochreiter} {and} \bibinfo{person}{J{\"u}rgen Schmidhuber}.} \bibinfo{year}{1997}\natexlab{}.
\newblock \showarticletitle{Long short-term memory}.
\newblock \bibinfo{journal}{\emph{Neural computation}} \bibinfo{volume}{9}, \bibinfo{number}{8} (\bibinfo{year}{1997}), \bibinfo{pages}{1735--1780}.
\newblock


\bibitem[Honbu and Yanai(2022)]%
        {honbu2022unseen-seg-3}
\bibfield{author}{\bibinfo{person}{Yuma Honbu} {and} \bibinfo{person}{Keiji Yanai}.} \bibinfo{year}{2022}\natexlab{}.
\newblock \showarticletitle{Unseen food segmentation}. In \bibinfo{booktitle}{\emph{Proceedings of the 2022 International Conference on Multimedia Retrieval}}. \bibinfo{pages}{19--23}.
\newblock


\bibitem[J{\'e}quier and Tappy(1999)]%
        {jequier1999weight-energy-balance}
\bibfield{author}{\bibinfo{person}{Eric J{\'e}quier} {and} \bibinfo{person}{Luc Tappy}.} \bibinfo{year}{1999}\natexlab{}.
\newblock \showarticletitle{Regulation of body weight in humans}.
\newblock \bibinfo{journal}{\emph{Physiological reviews}} \bibinfo{volume}{79}, \bibinfo{number}{2} (\bibinfo{year}{1999}), \bibinfo{pages}{451--480}.
\newblock


\bibitem[Jiang et~al\mbox{.}(2019)]%
        {jiang2019classification-5}
\bibfield{author}{\bibinfo{person}{Shuqiang Jiang}, \bibinfo{person}{Weiqing Min}, \bibinfo{person}{Linhu Liu}, {and} \bibinfo{person}{Zhengdong Luo}.} \bibinfo{year}{2019}\natexlab{}.
\newblock \showarticletitle{Multi-scale multi-view deep feature aggregation for food recognition}.
\newblock \bibinfo{journal}{\emph{IEEE Transactions on Image Processing}}  \bibinfo{volume}{29} (\bibinfo{year}{2019}), \bibinfo{pages}{265--276}.
\newblock


\bibitem[Jiao et~al\mbox{.}(2024)]%
        {jiao2024rode}
\bibfield{author}{\bibinfo{person}{Pengkun Jiao}, \bibinfo{person}{Xinlan Wu}, \bibinfo{person}{Bin Zhu}, \bibinfo{person}{Jingjing Chen}, \bibinfo{person}{Chong-Wah Ngo}, {and} \bibinfo{person}{Yugang Jiang}.} \bibinfo{year}{2024}\natexlab{}.
\newblock \showarticletitle{RoDE: Linear Rectified Mixture of Diverse Experts for Food Large Multi-Modal Models}.
\newblock \bibinfo{journal}{\emph{arXiv preprint arXiv:2407.12730}} (\bibinfo{year}{2024}).
\newblock


\bibitem[Kingma and Ba(2014)]%
        {kingma2014adam}
\bibfield{author}{\bibinfo{person}{Diederik~P Kingma} {and} \bibinfo{person}{Jimmy Ba}.} \bibinfo{year}{2014}\natexlab{}.
\newblock \showarticletitle{Adam: A method for stochastic optimization}.
\newblock \bibinfo{journal}{\emph{arXiv preprint arXiv:1412.6980}} (\bibinfo{year}{2014}).
\newblock


\bibitem[Kiourt et~al\mbox{.}(2020)]%
        {kiourt2020food-classification-3}
\bibfield{author}{\bibinfo{person}{Chairi Kiourt}, \bibinfo{person}{George Pavlidis}, {and} \bibinfo{person}{Stella Markantonatou}.} \bibinfo{year}{2020}\natexlab{}.
\newblock \showarticletitle{Deep learning approaches in food recognition}.
\newblock \bibinfo{journal}{\emph{Machine Learning Paradigms: Advances in Deep Learning-based Technological Applications}} (\bibinfo{year}{2020}), \bibinfo{pages}{83--108}.
\newblock


\bibitem[Lan et~al\mbox{.}(2023)]%
        {lan2023foodsam-seg-4}
\bibfield{author}{\bibinfo{person}{Xing Lan}, \bibinfo{person}{Jiayi Lyu}, \bibinfo{person}{Hanyu Jiang}, \bibinfo{person}{Kun Dong}, \bibinfo{person}{Zehai Niu}, \bibinfo{person}{Yi Zhang}, {and} \bibinfo{person}{Jian Xue}.} \bibinfo{year}{2023}\natexlab{}.
\newblock \showarticletitle{Foodsam: Any food segmentation}.
\newblock \bibinfo{journal}{\emph{IEEE Transactions on Multimedia}} (\bibinfo{year}{2023}).
\newblock


\bibitem[Liu et~al\mbox{.}(2020)]%
        {liu2020food-ingredients-3}
\bibfield{author}{\bibinfo{person}{Chengxu Liu}, \bibinfo{person}{Yuanzhi Liang}, \bibinfo{person}{Yao Xue}, \bibinfo{person}{Xueming Qian}, {and} \bibinfo{person}{Jianlong Fu}.} \bibinfo{year}{2020}\natexlab{}.
\newblock \showarticletitle{Food and ingredient joint learning for fine-grained recognition}.
\newblock \bibinfo{journal}{\emph{IEEE transactions on circuits and Systems for Video Technology}} \bibinfo{volume}{31}, \bibinfo{number}{6} (\bibinfo{year}{2020}), \bibinfo{pages}{2480--2493}.
\newblock


\bibitem[Liu et~al\mbox{.}(2024)]%
        {liu2024canteen}
\bibfield{author}{\bibinfo{person}{Guoshan Liu}, \bibinfo{person}{Yang Jiao}, \bibinfo{person}{Jingjing Chen}, \bibinfo{person}{Bin Zhu}, {and} \bibinfo{person}{Yu-Gang Jiang}.} \bibinfo{year}{2024}\natexlab{}.
\newblock \showarticletitle{From Canteen Food to Daily Meals: Generalizing Food Recognition to More Practical Scenarios}.
\newblock \bibinfo{journal}{\emph{IEEE Transactions on Multimedia}} (\bibinfo{year}{2024}).
\newblock


\bibitem[Liu et~al\mbox{.}(2021)]%
        {liu2021pyraformer}
\bibfield{author}{\bibinfo{person}{Shizhan Liu}, \bibinfo{person}{Hang Yu}, \bibinfo{person}{Cong Liao}, \bibinfo{person}{Jianguo Li}, \bibinfo{person}{Weiyao Lin}, \bibinfo{person}{Alex~X Liu}, {and} \bibinfo{person}{Schahram Dustdar}.} \bibinfo{year}{2021}\natexlab{}.
\newblock \showarticletitle{Pyraformer: Low-complexity pyramidal attention for long-range time series modeling and forecasting}. In \bibinfo{booktitle}{\emph{International conference on learning representations}}.
\newblock


\bibitem[Liu et~al\mbox{.}(2023)]%
        {liu2023itransformer}
\bibfield{author}{\bibinfo{person}{Yong Liu}, \bibinfo{person}{Tengge Hu}, \bibinfo{person}{Haoran Zhang}, \bibinfo{person}{Haixu Wu}, \bibinfo{person}{Shiyu Wang}, \bibinfo{person}{Lintao Ma}, {and} \bibinfo{person}{Mingsheng Long}.} \bibinfo{year}{2023}\natexlab{}.
\newblock \showarticletitle{iTransformer: Inverted Transformers Are Effective for Time Series Forecasting}. In \bibinfo{booktitle}{\emph{The Twelfth International Conference on Learning Representations}}.
\newblock


\bibitem[Lo et~al\mbox{.}(2019a)]%
        {lo2019point2volume-2}
\bibfield{author}{\bibinfo{person}{Frank P-W Lo}, \bibinfo{person}{Yingnan Sun}, \bibinfo{person}{Jianing Qiu}, {and} \bibinfo{person}{Benny~PL Lo}.} \bibinfo{year}{2019}\natexlab{a}.
\newblock \showarticletitle{Point2volume: A vision-based dietary assessment approach using view synthesis}.
\newblock \bibinfo{journal}{\emph{IEEE Transactions on Industrial Informatics}} \bibinfo{volume}{16}, \bibinfo{number}{1} (\bibinfo{year}{2019}), \bibinfo{pages}{577--586}.
\newblock


\bibitem[Lo et~al\mbox{.}(2019b)]%
        {lo2019point2volume-point-cloud-2}
\bibfield{author}{\bibinfo{person}{Frank P-W Lo}, \bibinfo{person}{Yingnan Sun}, \bibinfo{person}{Jianing Qiu}, {and} \bibinfo{person}{Benny~PL Lo}.} \bibinfo{year}{2019}\natexlab{b}.
\newblock \showarticletitle{Point2volume: A vision-based dietary assessment approach using view synthesis}.
\newblock \bibinfo{journal}{\emph{IEEE Transactions on Industrial Informatics}} \bibinfo{volume}{16}, \bibinfo{number}{1} (\bibinfo{year}{2019}), \bibinfo{pages}{577--586}.
\newblock


\bibitem[Makhsous et~al\mbox{.}(2019)]%
        {makhsous2019novelvolume-3}
\bibfield{author}{\bibinfo{person}{Sepehr Makhsous}, \bibinfo{person}{Hashem~M Mohammad}, \bibinfo{person}{Jeannette~M Schenk}, \bibinfo{person}{Alexander~V Mamishev}, {and} \bibinfo{person}{Alan~R Kristal}.} \bibinfo{year}{2019}\natexlab{}.
\newblock \showarticletitle{A novel mobile structured light system in food 3D reconstruction and volume estimation}.
\newblock \bibinfo{journal}{\emph{Sensors}} \bibinfo{volume}{19}, \bibinfo{number}{3} (\bibinfo{year}{2019}), \bibinfo{pages}{564}.
\newblock


\bibitem[Martinel et~al\mbox{.}(2018)]%
        {martinel2018classification-2}
\bibfield{author}{\bibinfo{person}{Niki Martinel}, \bibinfo{person}{Gian~Luca Foresti}, {and} \bibinfo{person}{Christian Micheloni}.} \bibinfo{year}{2018}\natexlab{}.
\newblock \showarticletitle{Wide-slice residual networks for food recognition}. In \bibinfo{booktitle}{\emph{2018 IEEE Winter conference on applications of computer vision (WACV)}}. IEEE, \bibinfo{pages}{567--576}.
\newblock


\bibitem[Meyers et~al\mbox{.}(2015)]%
        {meyers2015im2calories}
\bibfield{author}{\bibinfo{person}{Austin Meyers}, \bibinfo{person}{Nick Johnston}, \bibinfo{person}{Vivek Rathod}, \bibinfo{person}{Anoop Korattikara}, \bibinfo{person}{Alex Gorban}, \bibinfo{person}{Nathan Silberman}, \bibinfo{person}{Sergio Guadarrama}, \bibinfo{person}{George Papandreou}, \bibinfo{person}{Jonathan Huang}, {and} \bibinfo{person}{Kevin~P Murphy}.} \bibinfo{year}{2015}\natexlab{}.
\newblock \showarticletitle{Im2Calories: towards an automated mobile vision food diary}. In \bibinfo{booktitle}{\emph{Proceedings of the IEEE international conference on computer vision}}. \bibinfo{pages}{1233--1241}.
\newblock


\bibitem[Min et~al\mbox{.}(2019)]%
        {min2019survey}
\bibfield{author}{\bibinfo{person}{Weiqing Min}, \bibinfo{person}{Shuqiang Jiang}, \bibinfo{person}{Linhu Liu}, \bibinfo{person}{Yong Rui}, {and} \bibinfo{person}{Ramesh Jain}.} \bibinfo{year}{2019}\natexlab{}.
\newblock \showarticletitle{A survey on food computing}.
\newblock \bibinfo{journal}{\emph{ACM Computing Surveys (CSUR)}} \bibinfo{volume}{52}, \bibinfo{number}{5} (\bibinfo{year}{2019}), \bibinfo{pages}{1--36}.
\newblock


\bibitem[Min et~al\mbox{.}(2023)]%
        {min2023food2k}
\bibfield{author}{\bibinfo{person}{Weiqing Min}, \bibinfo{person}{Zhiling Wang}, \bibinfo{person}{Yuxin Liu}, \bibinfo{person}{Mengjiang Luo}, \bibinfo{person}{Liping Kang}, \bibinfo{person}{Xiaoming Wei}, \bibinfo{person}{Xiaolin Wei}, {and} \bibinfo{person}{Shuqiang Jiang}.} \bibinfo{year}{2023}\natexlab{}.
\newblock \showarticletitle{Large scale visual food recognition}.
\newblock \bibinfo{journal}{\emph{IEEE Transactions on Pattern Analysis and Machine Intelligence}} (\bibinfo{year}{2023}).
\newblock


\bibitem[Naritomi and Yanai(2021)]%
        {naritomi2021mesh}
\bibfield{author}{\bibinfo{person}{Shu Naritomi} {and} \bibinfo{person}{Keiji Yanai}.} \bibinfo{year}{2021}\natexlab{}.
\newblock \showarticletitle{3D Mesh Reconstruction of Foods from a Single Image}. In \bibinfo{booktitle}{\emph{Proceedings of the 3rd Workshop on AIxFood}}. \bibinfo{pages}{7--11}.
\newblock


\bibitem[Nie et~al\mbox{.}(2023)]%
        {Yuqietal-2023-PatchTST}
\bibfield{author}{\bibinfo{person}{Yuqi Nie}, \bibinfo{person}{Nam H.~Nguyen}, \bibinfo{person}{Phanwadee Sinthong}, {and} \bibinfo{person}{Jayant Kalagnanam}.} \bibinfo{year}{2023}\natexlab{}.
\newblock \showarticletitle{A Time Series is Worth 64 Words: Long-term Forecasting with Transformers}. In \bibinfo{booktitle}{\emph{International Conference on Learning Representations}}.
\newblock


\bibitem[Pan et~al\mbox{.}(2020)]%
        {pan2020chefganfood-img-generation-1}
\bibfield{author}{\bibinfo{person}{Siyuan Pan}, \bibinfo{person}{Ling Dai}, \bibinfo{person}{Xuhong Hou}, \bibinfo{person}{Huating Li}, {and} \bibinfo{person}{Bin Sheng}.} \bibinfo{year}{2020}\natexlab{}.
\newblock \showarticletitle{ChefGAN: Food image generation from recipes}. In \bibinfo{booktitle}{\emph{Proceedings of the 28th ACM International Conference on Multimedia}}. \bibinfo{pages}{4244--4252}.
\newblock


\bibitem[Radford et~al\mbox{.}(2021)]%
        {radford2021learningCLIP}
\bibfield{author}{\bibinfo{person}{Alec Radford}, \bibinfo{person}{Jong~Wook Kim}, \bibinfo{person}{Chris Hallacy}, \bibinfo{person}{Aditya Ramesh}, \bibinfo{person}{Gabriel Goh}, \bibinfo{person}{Sandhini Agarwal}, \bibinfo{person}{Girish Sastry}, \bibinfo{person}{Amanda Askell}, \bibinfo{person}{Pamela Mishkin}, \bibinfo{person}{Jack Clark}, {et~al\mbox{.}}} \bibinfo{year}{2021}\natexlab{}.
\newblock \showarticletitle{Learning transferable visual models from natural language supervision}. In \bibinfo{booktitle}{\emph{International conference on machine learning}}. PMLR, \bibinfo{pages}{8748--8763}.
\newblock


\bibitem[Salinas et~al\mbox{.}(2020)]%
        {salinas2020deepar}
\bibfield{author}{\bibinfo{person}{David Salinas}, \bibinfo{person}{Valentin Flunkert}, \bibinfo{person}{Jan Gasthaus}, {and} \bibinfo{person}{Tim Januschowski}.} \bibinfo{year}{2020}\natexlab{}.
\newblock \showarticletitle{DeepAR: Probabilistic forecasting with autoregressive recurrent networks}.
\newblock \bibinfo{journal}{\emph{International journal of forecasting}} \bibinfo{volume}{36}, \bibinfo{number}{3} (\bibinfo{year}{2020}), \bibinfo{pages}{1181--1191}.
\newblock


\bibitem[Salvador et~al\mbox{.}(2019)]%
        {salvador2019inversecooking}
\bibfield{author}{\bibinfo{person}{Amaia Salvador}, \bibinfo{person}{Michal Drozdzal}, \bibinfo{person}{Xavier Gir{\'o}-i Nieto}, {and} \bibinfo{person}{Adriana Romero}.} \bibinfo{year}{2019}\natexlab{}.
\newblock \showarticletitle{Inverse cooking: Recipe generation from food images}. In \bibinfo{booktitle}{\emph{Proceedings of the IEEE/CVF Conference on Computer Vision and Pattern Recognition}}. \bibinfo{pages}{10453--10462}.
\newblock


\bibitem[Salvador et~al\mbox{.}(2021)]%
        {salvador2021revampingretrieval-3}
\bibfield{author}{\bibinfo{person}{Amaia Salvador}, \bibinfo{person}{Erhan Gundogdu}, \bibinfo{person}{Loris Bazzani}, {and} \bibinfo{person}{Michael Donoser}.} \bibinfo{year}{2021}\natexlab{}.
\newblock \showarticletitle{Revamping cross-modal recipe retrieval with hierarchical transformers and self-supervised learning}. In \bibinfo{booktitle}{\emph{Proceedings of the IEEE/CVF Conference on Computer Vision and Pattern Recognition}}. \bibinfo{pages}{15475--15484}.
\newblock


\bibitem[Salvador et~al\mbox{.}(2017)]%
        {salvador2017recipe1M}
\bibfield{author}{\bibinfo{person}{Amaia Salvador}, \bibinfo{person}{Nicholas Hynes}, \bibinfo{person}{Yusuf Aytar}, \bibinfo{person}{Javier Marin}, \bibinfo{person}{Ferda Ofli}, \bibinfo{person}{Ingmar Weber}, {and} \bibinfo{person}{Antonio Torralba}.} \bibinfo{year}{2017}\natexlab{}.
\newblock \showarticletitle{Learning cross-modal embeddings for cooking recipes and food images}. In \bibinfo{booktitle}{\emph{Proceedings of the IEEE conference on computer vision and pattern recognition}}. \bibinfo{pages}{3020--3028}.
\newblock


\bibitem[Shukor et~al\mbox{.}(2022)]%
        {shukor2022transformerretrieval-4}
\bibfield{author}{\bibinfo{person}{Mustafa Shukor}, \bibinfo{person}{Guillaume Couairon}, \bibinfo{person}{Asya Grechka}, {and} \bibinfo{person}{Matthieu Cord}.} \bibinfo{year}{2022}\natexlab{}.
\newblock \showarticletitle{Transformer decoders with multimodal regularization for cross-modal food retrieval}. In \bibinfo{booktitle}{\emph{Proceedings of the IEEE/CVF Conference on Computer Vision and Pattern Recognition}}. \bibinfo{pages}{4567--4578}.
\newblock


\bibitem[Song et~al\mbox{.}(2024)]%
        {song2024}
\bibfield{author}{\bibinfo{person}{Fangzhou Song}, \bibinfo{person}{Bin Zhu}, \bibinfo{person}{Yanbin Hao}, {and} \bibinfo{person}{Shuo Wang}.} \bibinfo{year}{2024}\natexlab{}.
\newblock \showarticletitle{Enhancing Recipe Retrieval with Foundation Models: A Data Augmentation Perspective}. In \bibinfo{booktitle}{\emph{European Conference on Computer Vision}}.
\newblock


\bibitem[Suzuki et~al\mbox{.}(2020)]%
        {suzuki2020pointvolume-1}
\bibfield{author}{\bibinfo{person}{Takuo Suzuki}, \bibinfo{person}{Kana Futatsuishi}, \bibinfo{person}{Kana Yokoyama}, {and} \bibinfo{person}{Nobuko Amaki}.} \bibinfo{year}{2020}\natexlab{}.
\newblock \showarticletitle{Point cloud processing method for food volume estimation based on dish space}. In \bibinfo{booktitle}{\emph{2020 42nd Annual International Conference of the IEEE Engineering in Medicine \& Biology Society (EMBC)}}. IEEE, \bibinfo{pages}{5665--5668}.
\newblock


\bibitem[Tai et~al\mbox{.}(2023)]%
        {tai2023nutritionverse}
\bibfield{author}{\bibinfo{person}{Chi-en~Amy Tai}, \bibinfo{person}{Matthew Keller}, \bibinfo{person}{Saeejith Nair}, \bibinfo{person}{Yuhao Chen}, \bibinfo{person}{Yifan Wu}, \bibinfo{person}{Olivia Markham}, \bibinfo{person}{Krish Parmar}, \bibinfo{person}{Pengcheng Xi}, \bibinfo{person}{Heather Keller}, \bibinfo{person}{Sharon Kirkpatrick}, {et~al\mbox{.}}} \bibinfo{year}{2023}\natexlab{}.
\newblock \showarticletitle{NutritionVerse: Empirical Study of Various Dietary Intake Estimation Approaches}. In \bibinfo{booktitle}{\emph{Proceedings of the 8th International Workshop on Multimedia Assisted Dietary Management}}. \bibinfo{pages}{11--19}.
\newblock


\bibitem[Taylor and Letham(2018)]%
        {taylor2018Prophet}
\bibfield{author}{\bibinfo{person}{Sean~J Taylor} {and} \bibinfo{person}{Benjamin Letham}.} \bibinfo{year}{2018}\natexlab{}.
\newblock \showarticletitle{Forecasting at scale}.
\newblock \bibinfo{journal}{\emph{The American Statistician}} \bibinfo{volume}{72}, \bibinfo{number}{1} (\bibinfo{year}{2018}), \bibinfo{pages}{37--45}.
\newblock


\bibitem[Thames et~al\mbox{.}(2021)]%
        {thames2021nutrition5k}
\bibfield{author}{\bibinfo{person}{Quin Thames}, \bibinfo{person}{Arjun Karpur}, \bibinfo{person}{Wade Norris}, \bibinfo{person}{Fangting Xia}, \bibinfo{person}{Liviu Panait}, \bibinfo{person}{Tobias Weyand}, {and} \bibinfo{person}{Jack Sim}.} \bibinfo{year}{2021}\natexlab{}.
\newblock \showarticletitle{Nutrition5k: Towards automatic nutritional understanding of generic food}. In \bibinfo{booktitle}{\emph{Proceedings of the IEEE/CVF conference on computer vision and pattern recognition}}. \bibinfo{pages}{8903--8911}.
\newblock


\bibitem[Wang et~al\mbox{.}(2019)]%
        {wang2019retrieval-1}
\bibfield{author}{\bibinfo{person}{Hao Wang}, \bibinfo{person}{Doyen Sahoo}, \bibinfo{person}{Chenghao Liu}, \bibinfo{person}{Ee-peng Lim}, {and} \bibinfo{person}{Steven~CH Hoi}.} \bibinfo{year}{2019}\natexlab{}.
\newblock \showarticletitle{Learning cross-modal embeddings with adversarial networks for cooking recipes and food images}. In \bibinfo{booktitle}{\emph{Proceedings of the IEEE/CVF conference on computer vision and pattern recognition}}. \bibinfo{pages}{11572--11581}.
\newblock


\bibitem[Wang et~al\mbox{.}(2021)]%
        {wang2021crossretrieval-5}
\bibfield{author}{\bibinfo{person}{Hao Wang}, \bibinfo{person}{Doyen Sahoo}, \bibinfo{person}{Chenghao Liu}, \bibinfo{person}{Ke Shu}, \bibinfo{person}{Palakorn Achananuparp}, \bibinfo{person}{Ee-peng Lim}, {and} \bibinfo{person}{Steven~CH Hoi}.} \bibinfo{year}{2021}\natexlab{}.
\newblock \showarticletitle{Cross-modal food retrieval: learning a joint embedding of food images and recipes with semantic consistency and attention mechanism}.
\newblock \bibinfo{journal}{\emph{IEEE Transactions on Multimedia}}  \bibinfo{volume}{24} (\bibinfo{year}{2021}), \bibinfo{pages}{2515--2525}.
\newblock


\bibitem[Wang and Zhang(2020)]%
        {wang2020long-term-error}
\bibfield{author}{\bibinfo{person}{XiaoFeng Wang} {and} \bibinfo{person}{Ying Zhang}.} \bibinfo{year}{2020}\natexlab{}.
\newblock \showarticletitle{Multi-step-ahead time series prediction method with stacking LSTM neural network}. In \bibinfo{booktitle}{\emph{2020 3rd International Conference on Artificial Intelligence and Big Data (ICAIBD)}}. IEEE, \bibinfo{pages}{51--55}.
\newblock


\bibitem[Wu et~al\mbox{.}(2021b)]%
        {wu2021autoformer}
\bibfield{author}{\bibinfo{person}{Haixu Wu}, \bibinfo{person}{Jiehui Xu}, \bibinfo{person}{Jianmin Wang}, {and} \bibinfo{person}{Mingsheng Long}.} \bibinfo{year}{2021}\natexlab{b}.
\newblock \showarticletitle{Autoformer: Decomposition transformers with auto-correlation for long-term series forecasting}.
\newblock \bibinfo{journal}{\emph{Advances in neural information processing systems}}  \bibinfo{volume}{34} (\bibinfo{year}{2021}), \bibinfo{pages}{22419--22430}.
\newblock


\bibitem[Wu et~al\mbox{.}(2021a)]%
        {wu2021large-seg-2}
\bibfield{author}{\bibinfo{person}{Xiongwei Wu}, \bibinfo{person}{Xin Fu}, \bibinfo{person}{Ying Liu}, \bibinfo{person}{Ee-Peng Lim}, \bibinfo{person}{Steven~CH Hoi}, {and} \bibinfo{person}{Qianru Sun}.} \bibinfo{year}{2021}\natexlab{a}.
\newblock \showarticletitle{A large-scale benchmark for food image segmentation}. In \bibinfo{booktitle}{\emph{Proceedings of the 29th ACM international conference on multimedia}}. \bibinfo{pages}{506--515}.
\newblock


\bibitem[Yin et~al\mbox{.}(2023)]%
        {yin2023foodlmm}
\bibfield{author}{\bibinfo{person}{Yuehao Yin}, \bibinfo{person}{Huiyan Qi}, \bibinfo{person}{Bin Zhu}, \bibinfo{person}{Jingjing Chen}, \bibinfo{person}{Yu-Gang Jiang}, {and} \bibinfo{person}{Chong-Wah Ngo}.} \bibinfo{year}{2023}\natexlab{}.
\newblock \showarticletitle{FoodLMM: A Versatile Food Assistant using Large Multi-modal Model}.
\newblock \bibinfo{journal}{\emph{arXiv preprint arXiv:2312.14991}} (\bibinfo{year}{2023}).
\newblock


\bibitem[Zeng et~al\mbox{.}(2023)]%
        {zeng2023NLinear}
\bibfield{author}{\bibinfo{person}{Ailing Zeng}, \bibinfo{person}{Muxi Chen}, \bibinfo{person}{Lei Zhang}, {and} \bibinfo{person}{Qiang Xu}.} \bibinfo{year}{2023}\natexlab{}.
\newblock \showarticletitle{Are transformers effective for time series forecasting?}. In \bibinfo{booktitle}{\emph{Proceedings of the AAAI conference on artificial intelligence}}, Vol.~\bibinfo{volume}{37}. \bibinfo{pages}{11121--11128}.
\newblock


\bibitem[Zhou et~al\mbox{.}(2021)]%
        {zhou2021informer}
\bibfield{author}{\bibinfo{person}{Haoyi Zhou}, \bibinfo{person}{Shanghang Zhang}, \bibinfo{person}{Jieqi Peng}, \bibinfo{person}{Shuai Zhang}, \bibinfo{person}{Jianxin Li}, \bibinfo{person}{Hui Xiong}, {and} \bibinfo{person}{Wancai Zhang}.} \bibinfo{year}{2021}\natexlab{}.
\newblock \showarticletitle{Informer: Beyond efficient transformer for long sequence time-series forecasting}. In \bibinfo{booktitle}{\emph{Proceedings of the AAAI conference on artificial intelligence}}, Vol.~\bibinfo{volume}{35}. \bibinfo{pages}{11106--11115}.
\newblock


\bibitem[Zhou et~al\mbox{.}(2024)]%
        {zhou2024food-detection-1}
\bibfield{author}{\bibinfo{person}{Pengfei Zhou}, \bibinfo{person}{Weiqing Min}, \bibinfo{person}{Jiajun Song}, \bibinfo{person}{Yang Zhang}, {and} \bibinfo{person}{Shuqiang Jiang}.} \bibinfo{year}{2024}\natexlab{}.
\newblock \showarticletitle{Synthesizing knowledge-enhanced features for real-world zero-shot food detection}.
\newblock \bibinfo{journal}{\emph{IEEE Transactions on Image Processing}} (\bibinfo{year}{2024}).
\newblock


\bibitem[Zhou et~al\mbox{.}(2022)]%
        {zhou2022fedformer}
\bibfield{author}{\bibinfo{person}{Tian Zhou}, \bibinfo{person}{Ziqing Ma}, \bibinfo{person}{Qingsong Wen}, \bibinfo{person}{Xue Wang}, \bibinfo{person}{Liang Sun}, {and} \bibinfo{person}{Rong Jin}.} \bibinfo{year}{2022}\natexlab{}.
\newblock \showarticletitle{Fedformer: Frequency enhanced decomposed transformer for long-term series forecasting}. In \bibinfo{booktitle}{\emph{International conference on machine learning}}. PMLR, \bibinfo{pages}{27268--27286}.
\newblock


\bibitem[Zhu and Ngo(2020)]%
        {zhu2020cookganfood-img-generation-2}
\bibfield{author}{\bibinfo{person}{Bin Zhu} {and} \bibinfo{person}{Chong-Wah Ngo}.} \bibinfo{year}{2020}\natexlab{}.
\newblock \showarticletitle{CookGAN: Causality based text-to-image synthesis}. In \bibinfo{booktitle}{\emph{Proceedings of the IEEE/CVF Conference on Computer Vision and Pattern Recognition}}. \bibinfo{pages}{5519--5527}.
\newblock


\bibitem[Zhu et~al\mbox{.}(2019)]%
        {zhu2019r2ganretrieval-6}
\bibfield{author}{\bibinfo{person}{Bin Zhu}, \bibinfo{person}{Chong-Wah Ngo}, \bibinfo{person}{Jingjing Chen}, {and} \bibinfo{person}{Yanbin Hao}.} \bibinfo{year}{2019}\natexlab{}.
\newblock \showarticletitle{R2gan: Cross-modal recipe retrieval with generative adversarial network}. In \bibinfo{booktitle}{\emph{Proceedings of the IEEE/CVF Conference on Computer Vision and Pattern Recognition}}. \bibinfo{pages}{11477--11486}.
\newblock


\bibitem[Zhu et~al\mbox{.}(2020)]%
        {zhu2020cross}
\bibfield{author}{\bibinfo{person}{Bin Zhu}, \bibinfo{person}{Chong-Wah Ngo}, {and} \bibinfo{person}{Jing-jing Chen}.} \bibinfo{year}{2020}\natexlab{}.
\newblock \showarticletitle{Cross-domain cross-modal food transfer}. In \bibinfo{booktitle}{\emph{Proceedings of the 28th ACM International Conference on Multimedia}}. \bibinfo{pages}{3762--3770}.
\newblock


\end{thebibliography}










\end{document}